\documentclass[journal,draftcls,comsoc,onecolumn,12pt]{IEEEtran}
\IEEEoverridecommandlockouts
\usepackage{graphics} % for pdf, bitmapped graphics files
\usepackage{epsfig} % for postscript graphics files
\usepackage{mathptmx} % assumes new font selection scheme installed
\usepackage{times} % assumes new font selection scheme installed
\usepackage{amsmath} % assumes amsmath package installed
\usepackage{amsthm}
\usepackage{amssymb}  % assumes amsmath package installed
\usepackage{commath}
\usepackage{mathtools}
\usepackage{comment}
\usepackage{tabularx}
\usepackage{mathtools,xparse}
\usepackage{newtxtext, newtxmath}
\usepackage[numbers,sort,square,compress]{natbib}
\usepackage{xcolor}
\usepackage{caption}
\DeclareMathOperator{\Tr}{Tr}

\newcommand{\sibi}[1]{{(\color{blue}#1})}
\newtheorem{theorem}{Theorem}
\title{
A Post-coder Feedback Approach to Overcome Training Asymmetry in MIMO-TDD
}
\author{Arun Kumar Miryala, Irina Merin Baby, Kumar Appaiah, Sibi Raj
  B. Pillai\\
  Department of Electrical Engineering\\
Indian Institute of Technology Bombay, Mumbai, India}

\def\bH{\mathbf{H}}
\def\bE{\mathbf{E}}
\def\bQ{\mathbf{Q}}
\def\bU{\mathbf{U}}
\def\bJ{\mathbf{J}}
\def\bV{\mathbf{V}}
\def\bu{\mathbf{u}}
\def\bv{\mathbf{v}}
\def\bx{\mathbf{x}}
\def\by{\mathbf{y}}
\def\bG{\mathbf{G}}
\def\bD{\mathbf{D}}

\def\bB{\mathbf{B}}
\def\bK{\mathbf{K}}
\def\bw{\mathbf{w}}
\def\bI{\mathbf{I}}
\def\eE{\mathbb E}
\def\tet{\eta}
\def\Cfcsi{\text{C}_{\text{full CSI}}}
\def\Rfcsi{\text{R}_{\text{full-CSI}}}

\def\eQ{(\hat{\bU}^{\dagger}\bU - \bI_r)}

\def\Prx{P_{\mathrm{rx}}}
\def\slloyd{\epsilon_{\mathrm{Lloyd}}^2}

\begin{document}

\maketitle
\thispagestyle{empty}
\pagestyle{empty}

\begin{abstract}
  Time Divison Duplex (TDD) wireless communication systems are
  inherently bidirectional, which facilitates exploiting channel
  reciprocity for pilot based channel estimation of both uplink and
  downlink. However, there exists a gross asymmetry in channel
  estimation complexity for the uplink and downlink, particularly for
  Multiple Input Multiple Output (MIMO) TDD systems. Usually, Base
  Stations (BS) with more antennas need to estimate fewer parameters
  from each antenna, whereas the estimation requirement is
  disproportionately higher at the User Equipment (UE). Unlike the UE,
  the BS has powerful hardware, computational resources and energy to
  accurately estimate and track channel profiles. To overcome this
  asymmetry, we propose a solution for MIMO-TDD downlink
  communication, wherein the post-coder part of the channel matrix is
  quantized at the BS, and is communicated to the UE via a low-rate
  channel. Using asymptotically tight lower bounds on the downlink
  achievable rates, we quantify the performance of the proposed
  scheme. Simulations reveal that a moderate number of quantization
  bits are sufficient to achieve rates close to the the link
  capacity. We further show that, when the BS has many more antennas
  at than the UE, the channel can be compensated by appropriate
  transmit domain precoding without post-coder knowledge at the UE.
  %For lower feedback rates, we show that
  %quantizing only the best singular values performs effectively.
\end{abstract}

%%%%%%%%%%%%%%%%%%%%%%%%%%%%%%%%%%%%%%%%%%%%%%%%%%%%%%%%%%%%%%%%%%%%%%%%%%%%%%%%
\section{Introduction}

The use of Multiple Input Multiple Output (MIMO) techniques has been pioneered as one of the key approaches to
enhance data rates of wireless systems. When the Base Station (BS) possesses a large number of
antennas, the use of time-division duplex (TDD) communication with
uplink training can significantly enhance downlink data rates and link
reliability \cite{larsson2014massive}. This concept has been explored in great detail and has been shown to work effectively in
practice. However, most considerations assume the receiving user
equipment (UE) to possess a single antenna. In such systems, the base station can effectively precode to compensate for channel effects, so that the receiver detection is significantly simplified. In particular, once the channel vector is known at the BS, knowing its
norm at the UE is sufficient to decode the downlink transmissions. In
other words, UE need not perform full fledged channel
estimation using downlink pilots, since this is a waste of
effort. However, the situation drastically changes when there are
multiple antennas at the UE, since estimating the channel norm at each
UE antenna is now insufficient for rate maximization. On the other
hand, training to estimate the full channel is not only wasteful, but
also an onerous task for a battery powered, low
form-factor UE.  Notice that  the computational and hardware
resources available at the BS allows it to near accurately estimate
and track the uplink channel, where only a few parameters per antenna
need to be estimated. The reverse link, through reciprocal in TDD,
requires estimating several coefficients per antenna, that too with
less resources, leading to a `curse of asymmetry' in pilot based
training. Moreover, different UEs may be equipped with varying number
of antenna and hardware, making a single-size-fits-all pilot training
infeasible and inaccurate.  In order to address this, we propose a
solution based on the feedback of partial channel state
information (CSI) measured at the base station to the UE, and demonstrate that this enhances the system throughput significantly, without compromising on simplicity.
Our feedback approach is particularly suitable for practical $r\times t$ MIMO systems where the number $t$ of antennas at BS ranges from $4$ to $12$, whereas the UE may be equipped with $2$ to $4$ antennas.

%\sibi{the paragraph here was a bit self-destructive, it should focus on why reference [2] is supporting our cause. I have taken it off for now}
\begin{comment}
In the context of massive MIMO, it has been shown that by increasing
the number of antennas at the base station, the asymptotic behaviour
of the wireless channel, along with effective processing techniques
can increase throughput significantly. In addition, massive MIMO
brings about additional benefits in terms of low-power components,
reduced latency, scalability and channel
hardening~\cite{larsson2014massive} which eliminates the need for
sharing singular values for the channel thereby reducing channel
state information (CSI) transmission overheads. Such systems generally
require the transmitter to know the CSI when the UE has a single
antenna. However, when UEs have multiple antennas, they also need to
possess some CSI for achieving the
capacity~\cite{telatar1999capacity}, viz. the knowledge of left
singular vectors of the channel matrix (which we refer to as the {\em
  post-coder}). Our work considers quantization and feedback of these
post-coders for achieving capacity with multi-antenna UEs.
\end{comment}

Typical wireless systems employ pre-processing of the transmitted symbols,
%using CSI feedback from the receiver to the transmitter.
%enabling the  preprocessing of the transmitted data.
leading to reduced receiver complexity as
well as higher data rates~\cite{love2008overview}. Acquiring receiver CSI places a huge burden on the UE in Frequency Division Duplex (FDD) links, particularly for
MIMO systems. The alternative TDD systems have scope for efficient workarounds,
%For example, a UE receiver with  a single antenna
%needs to know only the norm of the channel vector for achieving the
%capacity,
once the BS has acquired the full CSI vector of the reciprocal channel. It is of crucial importance to then
convey the channel singular values as well as the left
unitary matrix (called post-coder) of the downlink, under Singular Value Decomposition (SVD), to the UE.
%
%Moreover, achieving
%capacity \sibi{what capacity?} requires accurate knowledge of the CSI at the transmitter.
%In TDD systems, the reciprocity of the channel in both directions can be
%exploited to obviate the need to possess CSI at the single antenna
%UEs.
%For multi-antenna UEs, the post-coder matrix can be sent to the UE using
%a separate channel, akin
The low rate control channels present in the standards~\cite{donthi2011joint} can be exploited for such CSI feedback.
%These are typically low rate channels, with limited bit budget.
%It is now essential to analyze the loss of performance due to
%the quantization of post-coder matrix.
Notice that the feedback bit budget is
critical for block fading channels, where frequent channel updates using a small number of bits are required, while adaptive
feedback can reduce this load  for smoother channel variations. In any case, it is essential
to understand the loss of performance due to the quantization of CSI.
%of  post-coder matrix.

Let us focus on conveying the post-coder matrix to the UE. This requires efficient quantization of
unitary matrices. % that represents the post-coder.
Typical approaches to quantize unitary matrices  involve the use of vector
quantization on manifolds, such as the Grassmannian and Stiefel
manifolds. These approaches primarily exploit the algebraic structure
of the precoder to obtain quantization schemes (unit
vectors~\cite{love2003limited}, semi unitary
matrices~\cite{schwarz2014predictive} and unitary
matrices~\cite{choi2006interpolation}). Though these methods have
been shown  effective in theory, they involve
operations on high dimensional manifolds, often an overkill
%conveying the post-coder unitary matrix
due to the complex processing
requirements. On the other hand, it is known that unitary matrices can be efficiently decomposed into
independent parameters using Givens rotations and Householder
transformations~\cite{rohquantisation}. The utility of this decomposition is epitomized by their adoption
in several wireless standards~\cite{kim2015802}. Moreover, the parameters obtained in this
approach are independent and can be quantized using scalar
quantizers. Another advantage is that in slowly varying channels, the temporal evolution of these scalar parameters can be tracked using single-bit adaptive
quantizers~\cite{roh2004efficient}.

The main objective of the current paper is to propose efficient post-coder quantization schemes for TDD MIMO using Givens rotation, which not only fit the bit budget, but also significantly enhance the downlink achievable rates.
%
%To facilitate this assume that the channel singular values are
%In this paper, we propose efficient post-coder quantization schemes, and show that a
%significant enhancement in performance can be alluded to the post-coder
%feedback to the UE.
In order to characterize the quantizer performance, we perform an information theoretic analysis, assuming the availability of channel singular values and quantized post-coder CSI at the UE. We will show that rate gap between our scheme and the capacity with full CSI  vanishes exponentially in the number of feedback bits. Furthermore, numerical computations reveal that a small number of feedback bits
suffices to achieve rates close to the ergodic capacity using full CSI.
The approaches suggested here are also attractive due to the fact that several
standards already use the Givens rotations to quantize precoder
information~\cite{kim2015802}, and the enhancements proposed here can
be realistically adopted in practical systems.

For UEs having a few antennas, channel estimation and feedback requirements may behave differently as we add more and more antennas at the BS. More specifically, channel hardening will have a big impact in systems with hundreds of transmit antennas. However, to reap the benefits of such massive MIMO systems, without the receiver having access to the post-coder,  non-convenional techniques may be required. For example,
conventional water-filling based transmit power allocation approaches appear to perform worse in the absence of  full receiver CSI. For completeness, we also propose efficient precoding based communication schemes for massive MIMO systems without post-coder information at the receiver, and analyze the resulting rates.

The remaining part of this paper is organized as follows:
Section~\ref{sec:systemmodel} describes the system
model.  In Section~\ref{sec:feedback}, we propose an
efficient post-coder quantization scheme, and analytically
characterize the achievable rate performance as a function of
the number of feedback bits. Furthermore, we propose schemes for massive MIMO with very limited feedback of CSI from the BS to the UE.  Section~\ref{sec:simulations} numerically compares the performance for various MIMO configurations, and shows that
the suggested schemes are very efficient. Finally, Section~\ref{sec:conclusion} concludes the paper with some future directions.

\section{System model}
\label{sec:systemmodel}
\begin{figure}[htbp]
\centering
\includegraphics[width=1\textwidth]{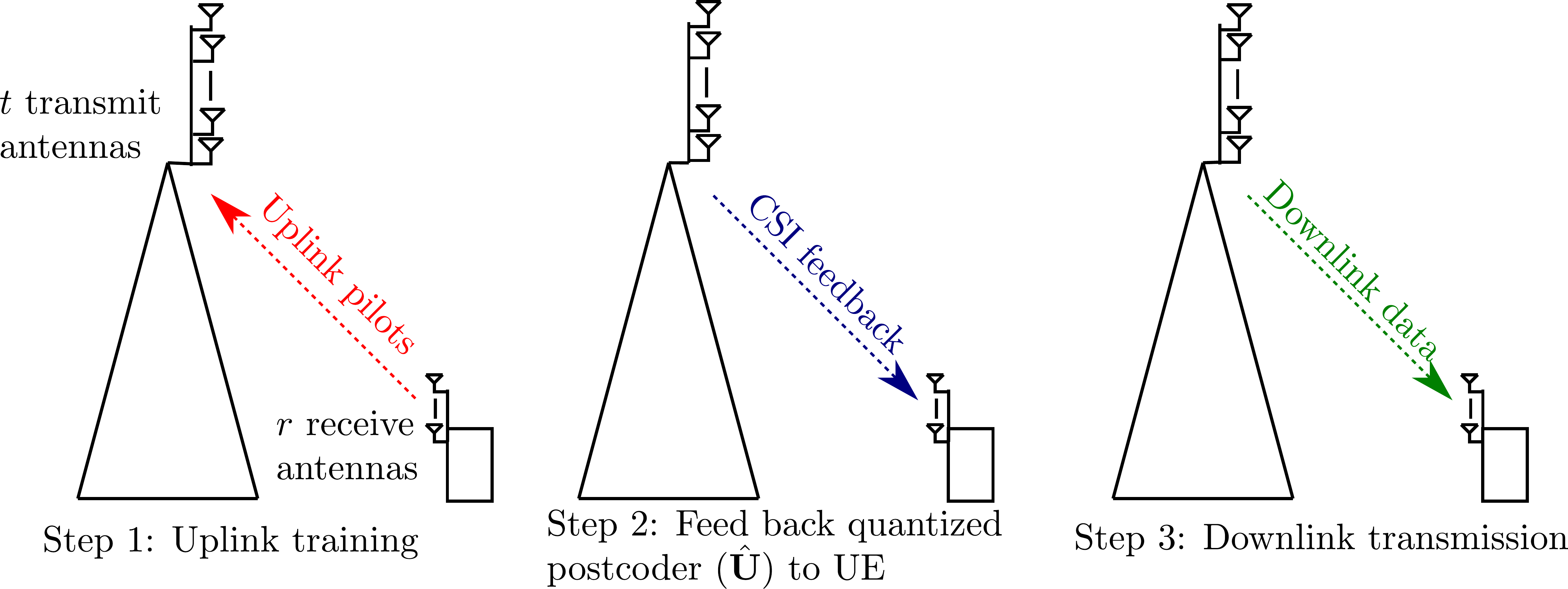}
\caption{\label{fig:situation}Training and feedback in the MIMO TDD
  system. The base station first learns the channel from uplink
  pilots (Step 1), and sends back just the compressed (as CSI feedback)
  post-coder to the UE (Step 2), which uses it to decode data (Step 3).}
\end{figure}
We consider a TDD-MIMO system with $t$ antennas at the transmitter and
$r$ antennas at the receiver. In this system, the channel is estimated
at the base station using uplink pilots, and due to channel
reciprocity, the downlink channel is assumed to possess the same
channel coefficients. We assume a block fading model with Rayleigh
distributed channel coefficients, wherein the MIMO channel is modeled
by a channel matrix $\bH \in\mathbb{C}^{r \times t}$ whose entries are
i.i.d. unit normal complex Gaussian random variables. That is, when
the input $\tilde{\bx} \in \mathbb{C}^{t}$ is sent by the transmitter
(base-station) to the receiver (UE), the received symbols
$\tilde{\by} \in \mathbb{C}^r$
%where $\tilde{\by}$ and $\tilde{\bx}$ are
%related as
are given by
\begin{equation}
\label{eq:firsteq}
\tilde{\by} = \bH\tilde{\bx} + \mathbf{\eta}.
\end{equation}
Here $\mathbf{\eta} \in \mathbb{C}^{r}$ is the additive white Gaussian
noise vector with i.i.d. entries, each having unit variance. We
assume that $t > r$, thus making the $\mbox{rank}(\bH)=r$ with high
probability. Due to reciprocity, the channel from the UE to the BS is
then ${\bH}^T$~\cite{larsson2014massive}. The singular value
decomposition (SVD) of the $\bH$ is given by
$\bH=\bU \Sigma \bV^{\dagger}$ where $\bU\in \mathbb{C}^{r\times r}$
and $\bV\in \mathbb{C}^{t\times r}$ are unitary matrices and
$\Sigma\in \mathbb{R}^{r\times r}$ contains the non-negative singular values
$\sigma_{1}\geq\cdots\geq\sigma_{r}$ of $\bH$. Also, an average
transmit power
constraint is imposed by stipulating $\eE[\tilde{\bx}^{\dagger}\tilde{\bx}]\leq P_{{T}}$, where the
expectation is over the transmitted codewords, across fading blocks.

As shown in Fig.~\ref{fig:situation}, the transmission within each
coherence interval occurs in three steps
\cite{larsson2014massive}. The first step involves uplink training for
channel estimation. Here, the pilot symbols
${\bf p}_1, {\bf p}_2, \ldots {\bf p}_{\tau}$ (all belonging to
$\mathbb{C}^{r}$), are sent over $\tau$ time instances. The signal
received at the base station when the UE transmits these pilots is described by
\begin{equation}
{\by}_{B,i} = \bH^T {\bf p}_i + \mathbf{\eta}_{B,i}\,,\,\,\text{ }i =  1,2, \ldots \tau.
\end{equation}
The channel can then be estimated at the base station using
${\by}_{B,i}, i =  1, \cdots, \tau$. We assume that the CSI obtained by
this approach is error-free, i.e. full CSI at Transmitter (CSIT).
If full CSI is also available at Reciever (CSIR), then the capacity of
this model is well known~\cite{telatar1999capacity, goldsmith2005wireless, TseViswanath05}.
In particular, $r$ parallel channels are effectively available in each fading block~\cite{telatar1999capacity}, and the  average transmit power is optimally divided over these parallel channels,  and also across fading blocks, to achieve the ergodic capacity.
%The base station may now use this
%information to optimally allocate power across the $r$ parallel
%channels obtained using the SVD, as is done
%for MIMO Gaussian channels \cite{telatar1999capacity}. To achieve this
%capacity, the UE also needs to possess knowledge of the channel, i.e.
Observe that once the BS knows $\bH =\bU \Sigma \bV^{\dagger}$, the
UE needs to know $\sigma_1, \sigma_2, \cdots, \sigma_r$ as well as $\bU$,
to achieve this capacity. The key question is how to communicate
the relevant information to the UE. Employing downlink pilots is one such option, but
can really be a burden to the UE. Instead, notice that TDD permits another natural feedback mechanism as follows.
%In
%particular, the BS can find the good downlink eigen modes and apply
%channel inversion on the corresponding singular values.  This allows
%the BS to further invert the effect of $\bU$ as well.
The BS can perform simple channel inversion using the pseudo-inverse of $\bH$,
and create low-rate parallel channels to convey the feedback bits.
While this inversion technique can be suboptimal from a capacity
perspective, the BS can now reliably convey the relevant channel
parameters in a small number of transmissions. Selective inversion, after
ignoring the low singular values of $\bH$, is another tangible technique. Alternately,  feedback can be done using the available downlink control channels as well.
%In any case, since the effect of $V^{\dagger}$ can be removed by an appropriate
%rotation at the BS, only $\Sigma$ and $\bU$ need to be conveyed using
%the feedback link.
Since $\bU$ has more parameters than $\Sigma$, we will focus on
feeding back a quantized version $\hat{\bU}$.
Once the post-coder is reconstructed at the receiver,
simple strategies like energy detection can be used to estimate the
effective singular values. For simplicity, we assume in our analysis
that $\Sigma$ is conveyed correctly to the receiver or known by other
techniques.

Quantization of $\bU$ to obtain $\hat{\bU}$
results in some loss of information, thereby causing the receiver to
not compensate for $\bU$ completely.
%Feeding back
%this information from the base station to the UE is possible using the control channels available in standards, or using other approaches.
On the other hand, since the quantized feedback necessitates some additional transmissions,
it is essential to minimize overheads while maximizing the data rate. We thus explore the
trade-off between the overheads for transmitting $\hat{\bU}$, and the
achievable downlink rates. To this end, we propose efficient quantization
schemes for $\bU$ under a specified bit constraint, and then attempt to maximize the resulting  achievable rate. Under full CSIT and the knowledge of $(\hat\bU,\Sigma)$ at the receiver, the maximum achievable rate can be computed by~\cite{TseViswanath05}
\begin{align} \label{eq:mut:inf}
\max_{p(\tilde{x}|\Sigma,\hat \bU,\bV)} I(\tilde{\bx};y|\Sigma, \hat\bU) \text{ s.t. } \eE \left[ \tilde{\bx}^{\dagger}\tilde{\bx} \right] \leq P_T.
\end{align}
Since full CSIR is not available, it is unclear
whether a  continuous valued distribution can maximize \eqref{eq:mut:inf},
let alone the Gaussian distribution. In spite of this difficulty, we will
show that appropriate Gaussian codebooks perform very well, and achieve
rates close to the capacity with full CSI, even while using a moderate number of quantization bits. Notice the difference between conventional fading models where the receiver is typically assumed to have
a better version of CSI than that available at the transmitter~\cite{TseViswanath05}, this adds to the novelty of our
analysis.
%We then investigate the quantization of both $\Sigma$ and $\bU$,
%under a fixed bit budget.
Our numerical results
identify good rules of thumb for effectively utilising the feedback bits.
%We will then numerically analyze the effects of quantizing both $\Sigma$
%and $\bU$ under a common bit budget constraint.

%\sibi{commented out something in the *.tex file}
%Since $\Sigma$ has $r$ non-zero entries, and
%$\bU$ is a $r\times r$ matrix, we consider the case where $\Sigma$ is
%directly conveyed (or is already known at the UE due to channel
%hardening \sibi{true for singular vectors as well}), whereas the unitary matrix $\bU$ needs to be quantized and
%fed back to the UE.

\section{CSI Feedback Scheme  and Achievable Rates}
\label{sec:feedback}
\subsection{Codebook to Compress Unitary Post-coders}\label{code_book}
To effectively feedback information about the matrix $\bU$ to the UE, a codebook
to compress unitary matrices is needed. While several codebook design
methods exist, we use the Givens rotation and Householder
transformation based decomposition to parameterize the unitary matrix,
and compress the parameters. This approach has the advantage that the
matrix is represented using scalar angle parameters that are all
independent and have well characterized probability
distributions~\cite{roh2007efficient}. The independence of these
parameters converts the compression problem to one where the optimal
quantizers for single dimensional random variables needs to be
found. These are well known, and thus, this approach is an effective
method to quantize the unitary matrices for feedback to the UE.

The Givens rotation and Householder transformations use rotation matrices
to null out off-diagonal elements of the unitary matrix. Since rotation matrices do not
change the norm of the matrix, the orthogonality of the unitary matrix $\bU$ is
preserved in the parametrization phase. First, the angles of each
element of the first column of the matrix are stored, and then the
entries of the column are made real. Subsequently, each of the
elements other than the first element are made zero using the
appropriate rotation matrix obtained from Givens rotations, and all
such angles are stored. These angles form the set of parameters needed
for reconstructing the $\bU$ matrix. Now, the above two steps are
repeated for all the columns in the matrix, making the resulting matrix to
be an identity matrix and the corresponding parameters are fed back for reconstruction at the receiver.

A unitary matrix $\bU\in C^{r \times r}  $ with orthonormal columns
can be decomposed as
 \begin{equation}\label{eq:pc:mat:rep}
   \bU=\left[\prod_{k=1}^{r-1}\bD_{k}(\phi_{k, 1}, \cdots, \phi_{k,
       k})\prod_{l=1}^{r-k}\bG_{r-l, r-l+1}(\theta_{k,
       l})\right]\bD_r(\phi_{r,1}, \ldots \phi_{r,r}),
 \end{equation}
where $\bD_{k}(\phi_{k, 1}, \cdots, \phi_{k, k})={\rm diag}({\bf 1}_{r-k},
\{ e^{j\phi_{k,j}}, 1\leq j \leq k \})$ and
\begin{equation*}\label{neqno:3}
  \bG_{p-1,p}(\theta)=\begin{bmatrix}
  {\bI_{p-2}}\cr
  {}&\cos\theta&-\sin\theta&{}\cr
  {}&\sin\theta&\cos\theta&{}\cr
  {}&{}&{}&\bI_{t-p}
  \end{bmatrix}.
\end{equation*}
We note that this is just the transposed representation of the unitary
matrix decomposition given in~\cite{roh2007efficient}. Here $\bI_{q}$ represents the $q\times q$ (block) identity matrix and
${\bf 1}_{r-k}$ represents a vector of $r-k$ ones. We remark that this
expansion is obtained by just taking the transposed form of the
expansion in~\cite{roh2007efficient}. The matrices $\bD_i$ and
$\bG_{j,k}$ contain parameters $\{(\phi_{k,i},\theta_{i,l})\}$ which,
after quantization, are fed back to the receiver. Reconstruction of
the matrix is a similar process, in which the operations are performed
in reverse. We note that the decomposition can also be performed by
performing the operations described above using the rows of the
matrix, as opposed to the columns, as we have done in the discussions
below.

The number of parameters required to be fed back can be reduced by
using the fact that the SVD is not unique. Since the angle information
in the left-singular matrix is required for the reconstruction of the
unitary matrix at the receiver, both the left and right singular
matrices are non-unique~\cite{trefethen1997numerical}, and can be
replaced with an equivalent pair of unitary matrices. Specifically,
for a matrix $\bH$ which is of size $r\times t$ and has rank $m$, we have:
\begin{equation}\label{neqno:4}
    \bH  = \sum^{m}_{i=1}\sigma_i \bu_i \bv_i^\dagger = \sum^{m}_{i=1}\sigma_i (e^{j\alpha_i})\bu_i (e^{j\alpha_i}\bv_i)^\dagger
\end{equation}
where $\bu_i, \bv_i$ represent the left and right singular vectors of
$\bH$ respectively, and the phase factors $\alpha_{i}$ can taken to be
any values in $(-\pi,\pi]$ without altering the SVD. Therefore, we can
choose $\alpha_i$ to make the first row of the unitary matrix $\bU$
real, thereby fixing the $m$ parameters to be zero and eliminating the
need to feed it back. For an $r \times t$ system having rank $r$, the
number of parameters needed to effectively feed back information about
the post-coder is $r^2-r$. This can be observed for a system having 2
antennas at UE as follows. First, any $2\times 2$ unitary matrix can
be uniquely described by the real parameters $\theta$ and
$\phi_1, \phi_2, \phi_3$ as follows~\cite{roh2007efficient}:
\begin{equation*}
  {\bU} =
  \underbrace{\begin{bmatrix}
      1& 0 &\\
      0&  e^{j \phi_1}&
  \end{bmatrix}}_{{\bU}_1}
  \underbrace{\begin{bmatrix}
      \cos  \theta & \sin  \theta &\\
      -\sin  \theta & \cos  \theta &
  \end{bmatrix}}_{{\bU}_2}
  \underbrace{\begin{bmatrix}
      e^{j \phi_2}& 0 &\\
      0&  e^{j \phi_3}&
  \end{bmatrix}}_{{\bU}_3}.
\end{equation*}
Notice that
\begin{align*}
H=\bU_1 \bU_2 \bU_3 \Sigma\, \bV^{\dagger} = \bU_1\bU_2 \Sigma\, \bU_3 \bV^{\dagger} = \bU_1 \bU_2 \,\Sigma \,(\bV\bU_3^{\dagger})^\dagger.
\end{align*}
We can observe that matrix $\bU_3$ is the diagonal matrix
$\bD_r(\cdot)$ in Equation~\eqref{eq:pc:mat:rep}. Thus $\bU_1\bU_2$ can be taken as the effective post-coder matrix, and $\bV\bU_3^{\dagger}$ becomes the precoder matrix, the latter having no effect on the received signal due to  precoding~\cite{telatar1999capacity}. Thus only $r(r-1)=2$ parameters need to be conveyed for $r=2$.
To simplify the parameter representation,
notations introduced here for angle parameters
(i.e., $\theta$ and $\phi_j$ $j = 1,2,3$)
will be used for any further discussions on
systems with only 2 antennas at the UE.
In general, the effective post-coder matrix to be communicated to the UE can be taken as
\begin{equation}\label{eq:pc:mat:rep:2}
   \bU=\left[\prod_{k=1}^{r-1}\bD_{k}(\phi_{k, 1}, \cdots, \phi_{k, k})\prod_{l=1}^{r-k}\bG_{r-l, r-l+1}(\theta_{k, l})\right]\bI.
 \end{equation}

%Therefore, if we account for the fact that an arbitrary transformation
%of the phases using a diagonal matrix involving angles %$\alpha_i$ values
%retains the SVD as equivalent, we can replace $\bU$ with
%\begin{equation}
%  {\bU\mathbf{D}} =
%  \begin{bmatrix}
%    1	& 0 &\\
%    0&  e^{j \phi_1}&
%  \end{bmatrix}
%  \begin{bmatrix}
%    \cos  \theta & \sin  \theta &\\
%    -\sin  \theta & \cos  \theta &
%  \end{bmatrix}
%  \begin{bmatrix}
%    e^{j \phi_2-\alpha_2}& 0 &\\
%    0&  e^{j \phi_3-\alpha_3}&
%    \end{bmatrix}.
%\end{equation}
%where
%$\mathbf{D} = \text{diag}(e^{-j\alpha_2}, %e^{-j\alpha_3})$.
%
%Choosing $\alpha_2 = \phi_2$, $\alpha_3 = \phi_3$ makes %the
%transformed ${\bU}_3$ matrix into an identity matrix, %thereby reducing
%2 angle parameters.
Subsequent to the
parameterization of the unitary matrix, we focus on effective
quantization of the angle parameters ($\phi_{i,j},\theta_{l,m}$). We use individual scalar quantizers for each independent parameter. %for compressing the matrix. Since the parameters
Since
$\phi_{i,j}\, {1\leq j \leq r}, {j\leq i \leq r}$ are distributed
uniformly in $(-\pi,\pi]$, uniform quantizers are suitable. The values
$\theta_{i,l} \ \forall \ i = \{ 1, \ldots, r-1 \},\ l = \{1,\ldots,
r-i\}$ have the probability density \def\stl{\epsilon^2_{\theta_l}}
\begin{equation}\label{eq:theta:pdf}
    p(\theta_{i,l})=2l\ (\sin\theta_{i,l})^{(2l-1)}\cos\theta_{i,l}\,\,,\text{ } 0\leq\theta_{i,l} < {\pi\over 2},
\end{equation}
which only depends on the index $l$. The optimal scalar quantizer
given by the Lloyd's algorithm~\cite{gersho2012vectorllyods} can now
be used to quantize each of these parameters. Let $\stl$ be the Mean
Squared Error (MSE) of the scalar quantizer for the parameter
$\theta_{i,l}$ in \eqref{eq:theta:pdf}.
%
%We next quantify the bound on the error in representing the post-coder
%given a bit budget for compression.
The MSE  between the post-coder
matrix $\bU$ and its quantization $\hat\bU$, denoted as
$\slloyd$,
can be computed as
%$\eE ||\bU - \hat\bU||_F^2 = \eE \Tr
%(\hat{\bU}^\dagger\bU-\bI_r)^\dagger(\hat{\bU}^\dagger\bU-\bI_r)$,
%Let us denote this MSE as
%$\slloyd$, where
\begin{align}\label{eq:var:lloyd}
\slloyd = \eE ||\bU - \hat\bU||_F^2 =  \mathbb{E}\left[\Tr
  (\hat{\bU}^\dagger\bU-\bI_r)^\dagger(\hat{\bU}^\dagger\bU-\bI_r)\right],
\end{align}
where $||{\bf A}||_F$ denotes the Frobenius norm.

%the mean squared quantization error.

\begin{theorem}
  \label{thm:general}
%  Let $\bU$ be the $r\times r$ unitary post-coder matrix, which can be represented using $\phi_{i,j}$ $j=1,\ldots,r$ $i = j,\ldots,r$ and $\theta_{i,j}$ $i=1,\ldots,r-1$ $j = 1,\ldots,r-i$ \sibi{better to say `as in (4)'}.
  For the post-coder matrix $\bU$ represented by
  \eqref{eq:pc:mat:rep}, let $\hat\bU$ be the unitary matrix
  reconstructed from the quantized values of $\phi_{i,j}$ and
  $\theta_{i,j}$ obtained as per the Lloyd's algorithm, with $b_1$ bits used to quantize each $\phi_{i,j}$.
  %and $b_2(l)$ bits to quantize $\theta_{i,l}$
  Then the quantization MSE
  %$\epsilon_{\mathrm{Lloyd}}^2 =
  %\mathbb{E}\left[\norm{\bU-\hat\bU}_{F}^2\right]$
  satisfies
  \begin{equation}\label{eq:upper:bnd}
  \slloyd \leq r(r-1)(1-\mathrm{sinc}(2^{-b_1}))+
  \sum_{l=1}^{r-1}2 (r-l)\, \stl.
  %
  %\left(\mathbb{E}\left[\sin\left(\frac{\theta_{k,l}-\hat\theta_{k,l}}{2}\right)^2\right]\right)
\end{equation}
%\begin{equation}\label{eq:tightbound}
 %   \epsilon_{\mathrm{Lloyd}}^2 \leq r(r-1)
 %   \left(1-\mathrm{sinc}(2^{-b_1})\right)+\sum_{l =
 %     1}^{r-1}\frac{\beta\left(\frac{2}{3},\frac{l+1}{3}\right)^{3}l(r
 %   - l)}{3(2^{2b_2(l)+3})}
%\end{equation}
%where $\beta(x, y) = \int_0^1 t^{x-t}(1 - t)^{y -
%  1}dt$ is the Euler integral. %   \begin{equation}\label{generic_experr}
  % \epsilon_{\mathrm{Lloyd}}^2 \leq
  % r(r-1)(1-\mathrm{sinc}(2^{-b_1}))+\sum_{l =
  %   1}^{r-1}\sqrt{1 - \frac{1}{2l}} \frac{(\pi / 2)^3(r - l)}{12 \times 2^{2b_2(l)}}.
  % \end{equation}
  %where $\hat{\theta}_{i,j}$ is the quantized value of $\theta_{i,j}$
  %in the $2^b$ sized codebook for $\theta_{i,j}$ obtained using the
  %Lloyd's algorithm, and
\end{theorem}
\begin{proof}
  The proof %for Equation~\eqref{generic_experr}
  can be found in Appendix~\ref{upper_experr}.
\end{proof}
%Note that the above accounts for the fact that $r$ parameters are not
%useful for representing the post-coder.

\begin{figure}[htbp]
\centering
\includegraphics[width=0.8\textwidth]{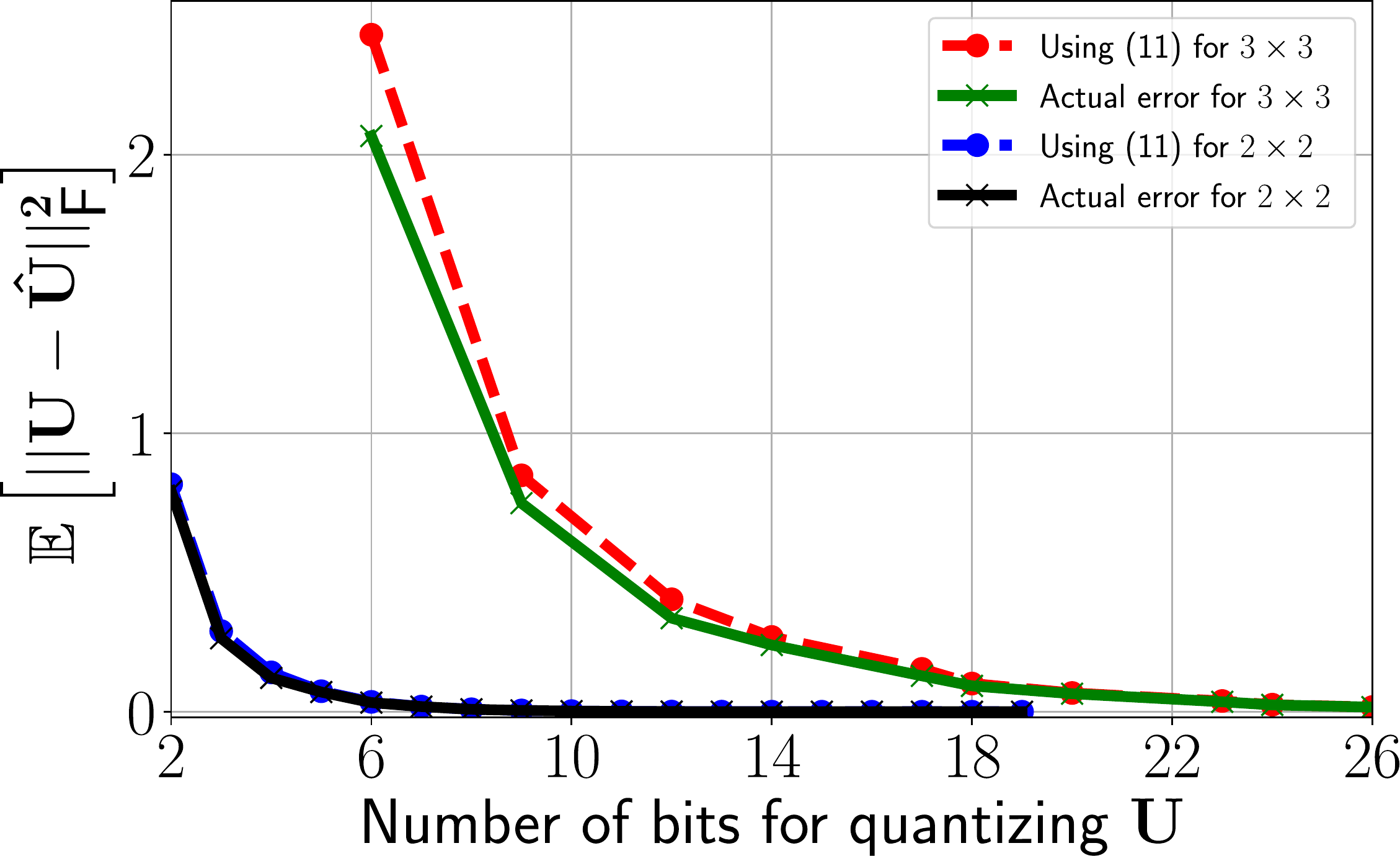}
\caption{\label{err_fig}Average post-coder quantization error: Actual Vs Equation~\eqref{eq:var:lloyd:bnd}.
%between  $\bU$ and the quantized post-coder $\hat{\bU}$.
\label{fig:lloyd:upper}}
\end{figure}

While the MSE $\stl$ in quantizing $\theta_{i,l}$
can be evaluated numerically using \eqref{eq:theta:pdf},
efficient closed form approximations are also available. In particular, the results in \cite{mselloyds} suggest that using $b_2$ bits for quantization will yield an MSE
\begin{equation*}
%\tilde{\theta}^2_{k,l}
\stl \approx \frac{1}{12(2^{2b_2})}\left(\int_{0}^{\frac{\pi}{2}} p(\theta_{k,l})^{\frac{1}{3}} \mathrm{d}\theta_{k,l}\right)^{3}.
\end{equation*}
Using \eqref{eq:theta:pdf} in the above expression, we get
\begin{equation}
  \label{mseequation}
%\tilde{\theta}^2_{k,l}
\stl
\approx
\frac{\beta\left(\frac{2}{3},\frac{l+1}{3}\right)^{3}l}{12(2^{2b_2+2})}\,,
\end{equation}
where $\beta(x,y), x\in\mathbb R^+, y\in \mathbb R^+$ is the beta function defined by~\cite{abramowitz1948handbook},
\begin{equation*}
  \beta(x,y) = \int_{0}^{1} t^{x-1}(1-t)^{y-1} \mathrm{d}t.
\end{equation*}
%for positive real numbers $x$ and $y$.
Using  %and~\eqref{expectheta},
 \eqref{eq:upper:bnd} and \eqref{mseequation} \begin{equation}\label{eq:var:lloyd:bnd}
  \epsilon_{\mathrm{Lloyd}}^2 \approx
  r(r-1)(1-\mathrm{sinc}(2^{-b_1}))+\sum_{l =
    1}^{r-1}\frac{\beta\left(\frac{2}{3},\frac{l+1}{3}\right)^{3}l(r
  - l)}{3(2^{2b_2(l)+3})},
\end{equation}
where $b_2(l)$ bits were used to quantize the parameter $\theta_{i,l}$ as per the  Lloyd's algorithm.  Clearly the Lloyd's algorithm achieves an exponential decay of MSE as more quantization bits are employed for each scalar parameter. This is also evident in \eqref{eq:var:lloyd:bnd}
%Theorem~\ref{thm:general} establishes that increasing the number of
%bits $b_1$ and $b_2$ allocated for quantizing the %parameters will reduce the
%quantization error as $O(2^{-2b_1}) + O(2^{-2b_2})$,  %As the number of bits used for
%quantization of each parameter ($b$) increases, since
as the functions $1-\mbox{sinc}(2^{-b})$ as well as
$\frac{\beta\left(\frac{2}{3},\frac{l+1}{3}\right)^{3}l(r -l)}{3(2^{2b+3})}$ decrease as $O(2^{-2b})$. While \eqref{eq:var:lloyd:bnd} was introduced as an approximate bound along the lines of \cite{mselloyds}, numerical results show that the RHS indeed gives a close upperbound to the MSE, as shown in Figure~\ref{fig:lloyd:upper}.

% \sibi{ let us change the labels of Fig.~2, the first one should be simply (i) Actual error for $2\times2$,
% (ii)Actual error for $3\times 3$,
% (iii) Using (11) for $2\times 2$,
% (iv) using (11) for $3\times 3$
% }
%
%\begin{figure}[htbp]
%\centering
%\includegraphics[width=0.8\textwidth]{./figures/bitsvserror_3ant}
%\caption{\label{err_fig_3}Average quantization error computed
%  numerically vs the average quantization error computed using Equation~\eqref{eq:var:lloyd:bnd}  between  $\bU$ and the quantized post-coder $\hat{\bU}$. }
%\end{figure}
%Thus, as $b_1$ and
%$b_2(l)$ increase, $\epsilon_{\mathrm{Lloyd}}^2$ vanishes
%exponentially to zero, thereby producing an accurate
%post-coder.
%
Thus, given a total bit budget of $b$ bits, we
can allocate $b_1$ bits for quantizing each $\phi_i$, and $b_2(l)$ bits to each $\theta_{i,l}$, in such a way that the MSE in \eqref{eq:var:lloyd:bnd} is minimized. By relaxing the integer constraints, straightforward solutions are possible for this minimization.

For particular configurations, the bound from Theorem~\ref{thm:general} can be made even more tight. For instance, the following theorem computes the exact MSE when $r=2$.
%Observe that
%$\epsilon_{\mathrm{Lloyd}}^2= \mathbb{E}\left[\Tr
%  (\hat{\bU}^\dagger\bU-\bI_r)^\dagger(\hat{\bU}^\dagger\bU-\bI_r)\right]$.
\begin{theorem}
  For a $2\times2$ MIMO system, let the unitary post-coder matrix $\bU$
  be parameterized by $\phi_1$ and
  $\theta$. Let each of these parameters be quantized using $b$ bits,
  and $\hat{\bU}$ be the post-coder reconstruction from the quantized
  values. Then,
  %the quantization error matrix, defined as
  %$ \mathbb{E}[(\hat{\bU}^\dagger\bU-\bI_r)^\dagger
  %(\hat{\bU}^\dagger\bU-\bI_r)]$ for such a system is given by
  \begin{equation}\label{eq:25}
%     \mathbb{E}[(\hat{\bU}^\dagger\bU-\bI_r)^\dagger
% (\hat{\bU}^\dagger\bU-\bI_r)]= 2\mathrm{sinc}(2^{-b})\begin{bmatrix}
%       1-\mathbb{E}[\cos(\theta-\hat\theta)] & 0 \\
%       0 & 1-\mathrm{sinc}(2^{-b}) \mathbb{E}[\cos(\theta-\hat\theta)]
%     \end{bmatrix}
\epsilon_{\mathrm{Lloyd}}^2 = 4 - 2\mathbb{E}\left[\cos(\theta - \hat{\theta})\right](1 + \mathrm{sinc}(2^{-b}))
% \epsilon_{\mathrm{Lloyd}}^2 = 2\mathrm{sinc}(2^{-b})\left(2 - (1 + \mathrm{sinc}(2^{-b}))
%   \mathbb{E}\left[\cos(\theta - \hat{\theta})
% \right]
% \right).
  \end{equation}
  where $\hat\theta$ is the quantized value of $\theta$.
\end{theorem}
\begin{proof}
  The proof %for Equation~\eqref{eq:25}
  is given  in Appendix~\ref{expecq}.
\end{proof}
%In Section~\ref{sec:simulations}, we will show results using this modified bound. \sibi{It will be good to highlight this in the simulations}

Our main objective now is to characterize the ergodic downlink rates under  quantized post-coder feedback to the receiver, in terms of the MSE $\slloyd$.

\subsection{Power Allocation and Ergodic Achievable Rates}
\label{sec:poweralloc}
To obtain the ergodic capacity of the system, it is necessary to allocate power in an optimal manner across blocks, based on the distribution of the channel and its current realization, while
maintaining a long term average power of $P_T$.  Notice that we assumed $r\leq t$,
i.e. the number of antennas at the UE $r$ is at most that at the BS, which has $t$ antennas.
%The average power available per time instant is $P$. Therefore, the total power that would be allocated in each time instant can be evaluated as described below.
%
Let us start by recollecting the optimal power control law when there is full CSI at the transmitter as well as receiver, this is based on the singular values
of the fading matrix~\cite{telatar1999capacity}.
The pdf of %unordered eigenvalues of $\bH^{\dagger}\bH$ of the channel matrix
the  unordered singular values of ${\bf H \bf H^{\dagger}}$ is given by~\cite{tulino2004random}:
\begin{equation*}
  g_{r,t}\left(\sigma\right) = \frac{1}{r}\sum_{k=0}^{r-1}\frac{k!}{(k+t-r)!}\left[L_{k}^{t-r}\left(\sigma\right)\right]^2\sigma^{t-r}e^{-\sigma},
\end{equation*}
%where $t$ and $r$ are the number of transmitter and receiver antennas
%respectively, as specified in Section~\ref{sec:systemmodel} above, and
where $L(\sigma)$ is the Laguerre polynomial given by
\begin{equation*}
  L_{k}^{n}(\sigma) = \frac{e^\sigma}{k!\sigma^n}\frac{d^k}{d\sigma^k}\left(e^{-\sigma}\sigma^{n+k}\right).
\end{equation*}
%Since the power to be allocated depends on the singular values of the
%instantaneous channel, we compute the amount of power that should be
%allocated to a particular singular vector based on its singular
%value.
The optimal power control is then given by
%This is done according to
the celebrated water-filling law~\cite{TseViswanath05}, which appropriately allocates
power over the $r$ parallel channels (one corresponding to each singular value) in each block, and across blocks as well. % To this end, let $\lambda$ represent the threshold on the
%singular value.
More specifically, a positive parameter $\lambda$ such that
\begin{equation}\label{expectationpower}
  %\mathbb{E}_{g(\sigma)}\left[\frac{1}{\lambda}-\frac{1}{\sigma^2}\right]^+ =
  \int_{\lambda}^{\infty} \left(\frac{1}{\lambda}-\frac{1}{\sigma^2}\right)g_{r,t}(\sigma)\,d\sigma = \frac{P_T}{r}
\end{equation}
is chosen, and $P(\sigma)= \max\{0,(\lambda^{-1}- \sigma^{-2})\}$ is the power allocated to the channel with singular value $\sigma$.
%Since $r \leq t$ by our assumption, let us take $\min(r,t) = r$.
The ergodic rate achievable with full CSI is then~\cite{TseViswanath05}
\begin{equation}\label{eq:ratefullcsi}
%  {C_{\text{full CSI}}} =
\Cfcsi= \log\left(
r \int_{\lambda}^{\infty}\frac{\sigma^2}{\lambda}  \,\, g_{r,t}(\sigma)\right)\,d\sigma.
\end{equation}
For future use, let us denote the average received power in each of the $r$ parallel channels as
\begin{align} \label{eq:pow:rx}
\Prx = \int_{\lambda}^{\infty} \left( \frac{\sigma^2}{\lambda} - 1\right)g_{r,t}(\sigma) d\sigma.
\end{align}
%
%\subsubsection{Lower bound on Achievable Rate}\label{sec:lowerbound}
%The discussions in this section so far
%~\ref{code_book} and
%Section~\ref{sec:poweralloc}
%permit us to
%
We now characterize the gap between  the achievable
rates obtained by the proposed quantization scheme,
and the capacity evaluated by \eqref{eq:mut:inf}.

\subsection{Gap to Ergodic Capacity} \label{sec:lowerbound}
It turns out that the proposed communication scheme under quantized feedback can achieve rates close
to \eqref{eq:ratefullcsi} itself, when a sufficient number of quantization bits are available.
The following theorem characterizes the rate gap to $\Cfcsi$ in terms of the quantization MSE.
\begin{theorem} \label{thm:lowerbound}
  For a MIMO channel matrix $\bH$ with i.i.d Rayleigh entries,
  the proposed  quantization scheme can achieve a rate of
  \begin{equation}\label{eq:rate:thm}
    \begin{split}
      \max_{p(x|\Sigma,\hat\bU)}I(\bx;\by |\Sigma,\hat\bU) & \geq %C_{\text{full CSI}}
      \Cfcsi - r\log_{2}\left(1+\frac{\Prx}{r}\epsilon_{\mathrm{Lloyd}}^2\right), \\
    \end{split}
  \end{equation}
  where
  $\slloyd$ and $P_{rx}$ are given in~\eqref{eq:var:lloyd} and~\eqref{eq:pow:rx} respectively, and $\Sigma = diag(\sigma_1,\cdots,\sigma_r)$.
\end{theorem}

\begin{proof}
  The proof is relegated to Appendix~\ref{lbproof}.
\end{proof}
While Theorem~\ref{thm:lowerbound} provides a convenient lower bound on the achievable rate for the downlink system, an astute reader might have observed that~\eqref{eq:tightlowerbound} can give even tighter bounds. In any case, since the MSE $\slloyd$
decreases exponentially fast  as more and more
bits are used for quantization, the average rates  quickly reach $\Cfcsi$.

%However, a closed for formula for $\slloyd$ in \eqref{eq:rate:thm} can enable the allocation of quantization bits to different parameters given in \eqref{eq:pc:mat:rep}, so as to minimize the overall MSE. Fortunately, an approximate MSE formula exists~\cite{}, which gives very close results for the scalar quantizers here.

\subsection{Massive MIMO Systems}
\label{sec:massivemimo}
While the techniques suggested so far enable us to achieve full CSI
capacity as finer quantizations of the post-coder become available,
they do not explicitly take into account the effect of channel
hardening present in massive MIMO systems. More specifically, for a
fixed number of UE antennas, the singular values and singular vectors
of the channel hardens or concentrates as the number $t$ of antennas
at the BS
increases~\cite{hochwald2004multiple,larsson2014massive}. While channel
hardening can lead to better quantizers having lower mean square
error, this may not always translate to an increase in achievable
rates. The major reason is that in order to reap the benefits of a significant boost in receive
signal strength achieved with massive MIMO, a post-coder matrix without
any mismatch at the receiver is required.
%Any mismatch may turn out
%harmful, placing more burden on the quantization scheme. Notice that
%our schemes will still approach
Thus achieving rates approaching the full CSI capacity will happen at the expense
of more quantization bits. However, if a constant gap to capacity is
admissible, then the burden on feedback can be considerably
reduced. In particular, the transmitter can precode  to avoid any
receiver mismatch, thereby avoiding the post-coder requirement. This is particularly appealing for massive MIMO, as the gap
to full CSI capacity is a small constant then.

%While we focused more on quantizing the post-coder matrix, typically,
%the singular values of the channel would also need to be known at the
%UE for optimal decodability from a rate perspective. In the next
%section, we will consider tradeoffs between quantizing $\bU$ and
%$\Sigma$. Notice that as the number of transmit antennas grows large,
%i.e.  $t \rightarrow \infty$, the singular values' distribution
%becomes more concentrated, as discussed
%in~\cite{hochwald2004multiple}. Therefore, the singular values do not
%need to be quantized, since the receiver can already possess
%information on the singular values using the hardening
%statistics~\cite{larsson2014massive} \sibi{this is also true for
%  post-coder, the reason by which we took it off}.

\def\bd{\mathbf d}
\def\als{\alpha({\Sigma})}
\def\sals{\sqrt{\alpha(\Sigma)}}

%Channel hardening enables us to directly compensate for the channel at
%the transmitter to some extent.
Our communication scheme works as
follows. Consider $r$ code-books of equal rate, where the entries of
each codebook are generated according to unit variance zero mean
Gaussian distribution. Thus $r$ transmitted data symbols at each
instant is denoted by the vector $\bd$. In order to convey $\bd$, the
transmitter sends
$\tilde x = \sals\, {\bf H}^\dagger({\bf H}{\bf H}^\dagger)^{-1} {\bf d}$ in
\eqref{eq:firsteq}, where $\sals$ is a positive real number that
depends on $\Sigma$. This translates to pre-multiplying the
transmitted signal by the scaled pseudoinverse of the channel. Notice
that this is not same as channel inversion, as the non-negative
scaling factor $\alpha(\Sigma)$ is a function of the singular values of
$\bH$, and this is crucial to our scheme. The receiver obtains
%
%Consider
%the following channel
%\begin{equation*}
%{\bf y} = {\bf H} {\bf x} + {\bf w}
%\end{equation*}
%Let the SVD of \({\bf H}\) be represented as \({\bf H} = {\bf
%U}\Sigma{\bf V}^H\). We let the actual transmit data be represented by
%the vector \({\bf d}\), with i.i.d. unit variance Gaussian entries. Then
%the actual transmission is \({\bf x} = \sqrt{\alpha}{\bf H}^H({\bf
%H}{\bf H}^H)^{-1} {\bf d}\), yielding
\begin{equation}
\label{eqn:precoded}
{\bf y} = \sals \, {\bf H}{\bf H}^\dagger({\bf H}{\bf H}^\dagger)^{-1}{\bf d}  + {\bf \eta}  = \sals \, {\bf d}  + {\bf \eta},
\end{equation}
where the choice of $\als$ ensures that the average power
constraint is satisfied. W.l.o.g assume that $\eta\sim \mathcal{NC}(0,\mathbb I)$.
%Now, we know that the power
%constraint dictates that
The constraint $ \mathbb{E}\left[\Tr({\bf x}{\bf
x}^\dagger)\right] \leq P_T$ will imply that
\begin{equation*}
\mathbb{E}\left[\alpha(\Sigma)\Tr(\Sigma^{-2})\right] \leq P_T.
\end{equation*}
The choice of $\als$ can now be made to maximize the resulting achievable rate.
Using the Lagrange multiplier $\lambda$, one can optimize by applying KKT conditions
on the unconstrained cost function
%To find the achievable rate, we can write the objective function \(R\) with
%the KKT parameter \(\lambda\) as follows:
\begin{equation}
R(\alpha(\Sigma), \lambda) = r\mathbb{E}\left[\log(1 + \alpha(\Sigma))\right] - \lambda\left(\mathbb{E}\left[\alpha(\Sigma)\, \Tr(\Sigma^{-2})\right] - P_T\right).
\end{equation}
Differentiating with respect to \(\alpha(\Sigma)\) for each channel realization, and
equating to zero,
we get
\begin{equation*}
\frac{r}{1 + \alpha(\Sigma)} = \lambda\, \Tr(\Sigma^{-2}).
\end{equation*}
Incorporating this back into the constraint and eliminating $\lambda$, the  rate expression
simplifies to
\begin{equation}
\label{eq:opt:zf}
R_{ZF} = r\, \eE \left[\log\left(\frac{P_T + \mathbb{E}[\Tr (\Sigma^{-2})]}{\Tr  (\Sigma^{-2})}\right)\right].
\end{equation}
%Thus, even if no feedback of the post-coder is made available to the
%UE, the above rate is achievable.
Remarkably, this rate is achieved without any CSI feedback to the
receiver at all.  We will show that this scheme performs reasonably
well, albeit with a small gap to capacity, when there are hundreds of
transmit antennas at the BS. We call this the ZF precoding approach.

For the particular case where the number of transmit and receive
antennas are the same, i.e. $r = t$, the above optimization does not
yield a bounded solution. In this situation, we selectively invert the
channel to transmit data only one stream of data that corresponds to
the largest singular value.

%However, due to the nature of the
%distribution of the singular values of ${\bf H}$, we note that the
%above optimization problem has no solution when the BS and UE have the
%same number of antennas, i.e. when $r = t$. In this case, post-coder
%feedback is the preferred strategy for %operating at a rate close to
%the capacity. \sibi{One can do selective inversion here, i.e. invert only the higher channels. Better to replace this comment by selective inversion}.

\section{Simulation results}
\label{sec:simulations}
We now present simulation results that characterize the downlink rate
based on the amount of feedback used to represent the post-coder $\bU$
at the receiver. We consider situations where the BS has several
antennas and a UE with $2$ or $3$ receive antennas. Quantized
post-coder feedback is beneficial here to approach the full CSI
capacity.  Finally, we comment on the performance of precoding in the
case of massive MIMO systems.

\begin{comment}
\sibi{I think it is better to write MIMO and Massive MIMO separately, lemme attempt}.
While we have presented general results for $t$ transmit antennas and $r$ receivers, the development targeted a BS
hosting up to a dozen antennas, and a UE with $2-4$ antennas. We will
call such models as MIMO systems, whereas a BS having one thousand or
so antennas will be termed as \emph{Massive MIMO} systems. Effects like
channel hardening becomes extremely important for the latter systems,
while it can be ignored for the former. In our simulations, we will
consider the two cases separately.
\end{comment}

\subsection{Quantization of unitary post-coder}
%We consider systems that have $10$ and $1000$
%transmit antennas at the base station and UEs that have $2$ or $3$
%antennas (i.e., $1000 \times 2$, $10 \times 2$, $1000 \times 3$ and
%$10 \times 3$ MIMO systems in the downlink case).
%
The first step in MIMO systems with quantized feedback is to have
codebooks for the unitary post-coder matrix. More specifically,
$\theta_{i,j}$  in \eqref{eq:pc:mat:rep} is quantized  as described in Section
\ref{code_book}, using the conventional Lloyd's
algorithm~\cite{gersho2012vectorllyods}. Since the distribution of
each $\phi_{i,j}$ is uniform, a uniform quantizer was
used for these parameters.
%
\begin{comment}
\subsection{Quantizer performance}
For a system that has $2$ antennas at the receiver, $\bU$ is a
$2\times 2$ matrix as given

\begin{equation*}
  {\bU} =
  \begin{bmatrix}
    1& 0 &\\
    0&  e^{j \phi_1}&
  \end{bmatrix}
  \begin{bmatrix}
    \cos  \theta & \sin  \theta &\\
    -\sin  \theta & \cos  \theta &
  \end{bmatrix}
  \begin{bmatrix}
    e^{j \phi_2}& 0 &\\
    0&  e^{j \phi_3}&
  \end{bmatrix}.
\end{equation*}

Since the SVD is not unique, as discussed in
Section~\ref{code_book}, this matrix can be represented using just $3$
parameters, viz. $\phi_2$, $\phi_3$ and $\theta$.
\end{comment}
Given a total feedback bit
budget, a bit allocation maximizing the rate bound of ~\eqref{eq:tightlowerbound} is used for each plot using our quantization scheme.  As an example,
for a $2\times 2$ post-coder matrix $\bU$, we need to quantize two
phase parameters, say $\theta$ and $\phi$. If $b$ bits are available,
we find the best combination of $b_1$ bits to quantize $\phi$ and
$b_2$ bits to quantize $\theta$ such that the maximum of \eqref{eq:tightlowerbound} is obtained and plotted in Fig. \ref{err_fig}. The reconstructed matrix obtained from the quantized
parameters, viz.  $\hat\phi$ and $\hat\theta$ is denoted as
$\hat{\bU}$. The quantization error is now measured as
$\mathbb{E}\left[\norm{\bU- \hat \bU }^{2}_{F}\right]$. %Fig.\ref{err_fig}, shows
The mean squared error in reconstruction is illustrated in
Fig. \ref{err_fig} for $2$ and $3$
antennas at the  UE, as a function of the total number of quantization bits
used to represent $\hat\bU$. Observe that about a dozen bits can make
the MSE sufficiently small for systems with $2$ antennas
at UE.

%\subsection{Lower bounds on achievable rate with quantized post-coders}

% \begin{figure}[!h]
%   \centering
%   \includegraphics[width=0.8\textwidth]{./figures/comb_lower_bound_without_sigma}
%   \caption{\label{Rate_fig} Lower bound on achievable rate when an equal
%     number of bits are allocated to all the parameters compared with bound when different
%     bits are allocated to  $\phi_3$ and $\theta$ for a $10\times2$ system.}
% \end{figure}

%One additional point to note is that we use a one-shot quantization
%based approach. However, a more appropriate approach would be to feed
%back the parameters of the quantization adaptively, using the approach
%suggested in \cite{roh2007efficient}. This way,
%$\epsilon_{\mathrm{Lloyd}}^2$ can be reduced even further as more fine
%grained quantization becomes available.

\subsection{Downlink Data Rates with post-coder feedback}
\label{sec:downlinkrates}
We now study how the quantization of $\bU$ impacts the achieved
rate. As discussed in Section~\ref{sec:massivemimo}, the post-coder
approach is expected to be beneficial when the number of antennas at the BS is
not very large, whereas the ZF precoding approach
%When the number of antennas at the BS increases, it is
%prudent to compensate for the channel completely at the transmitter
%using a zero-forcing precoder,
can be useful for massive MIMO systems where the
number of transmit antennas ($t$) far exceeds that at
the UE ($r$). However, observe that the optimization specified in~\eqref{eq:opt:zf} as such has no solution for $t = r$.

\begin{figure}[!h]
  \centering
  \includegraphics[width=\textwidth]{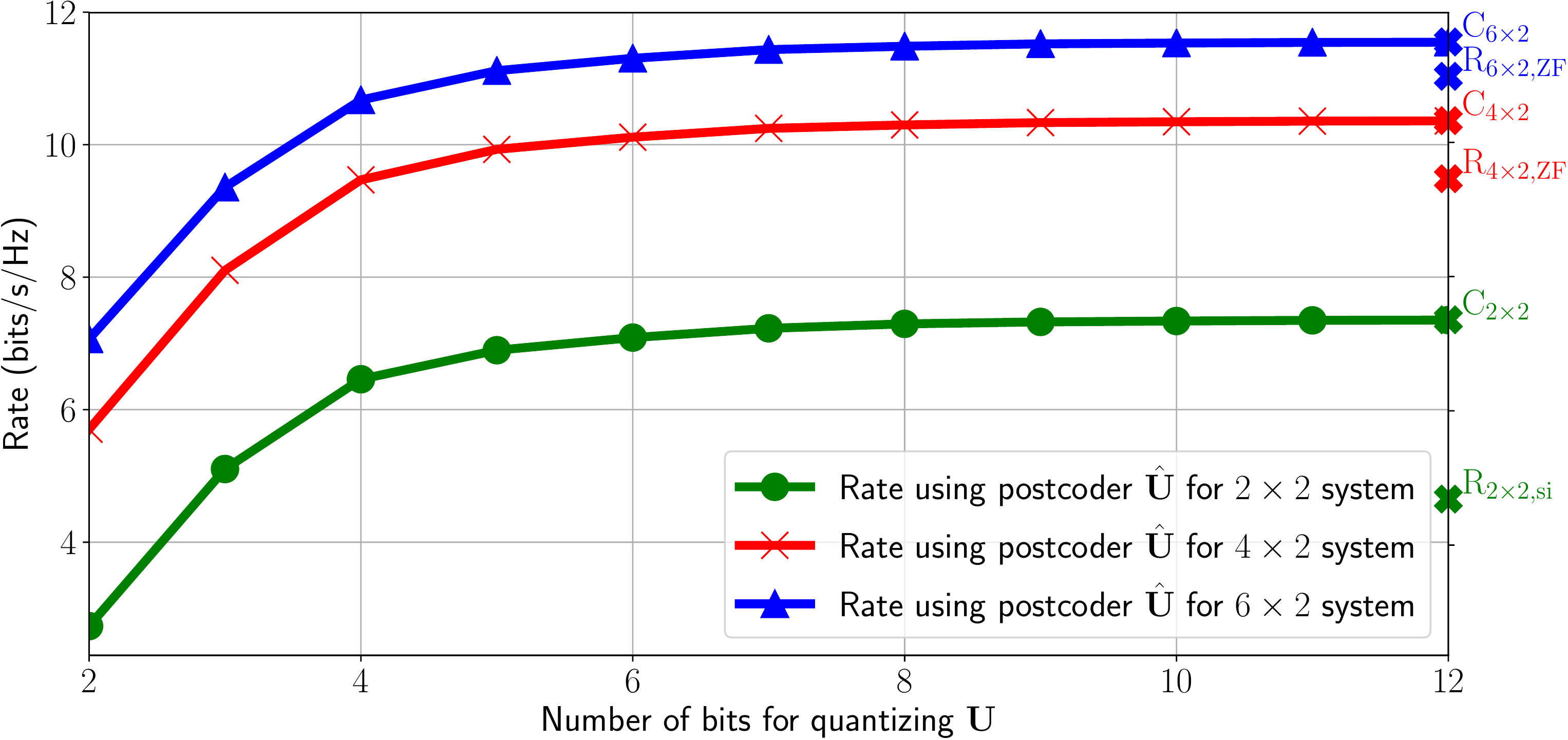}
  \caption{\label{Rate_fig_mul} Lower bound on achievable rate when optimal
    number of bits are allocated to  $\phi_3$ and $\theta$ compared with rate achieved
    using zero forcing precoding for $6\times 2$ $(\mathrm{R}_{6\times2,\mathrm{ZF}
    })$, $4\times2$ $(\mathrm{R}_{4\times2,\mathrm{ZF}})$ and $2\times2$ $(\mathrm{
      R}_{2\times2,\mathrm{ZF}})$ systems, where $\mathrm{C}_{6\times2}$, $\mathrm{
      C}_{4\times2}$ and $\mathrm{C}_{2\times2}$ are the respective system
    capacities at $10$ dB and $\mathrm{R}_{2\times2,\mathrm{si}}$ is the rate
    achieved using only the best Eigen mode for a $2\times2$ system.}
\end{figure}

\begin{figure}[!h]
\centering
\includegraphics[width=\textwidth]{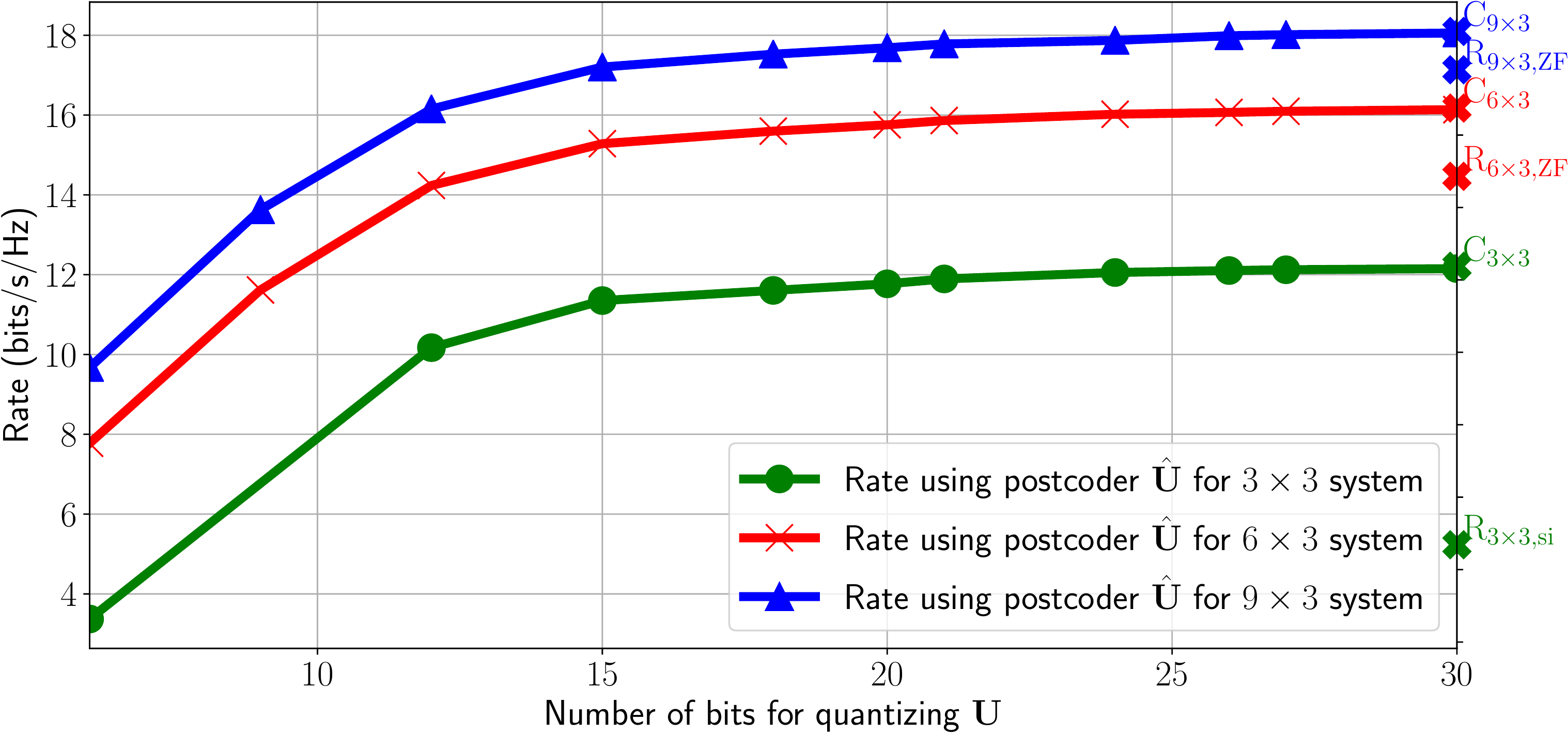}
\caption{\label{lower_bound_mul_ant_3ue} Lower bound on achievable
  rate when optimal number of bits are allocated to $\phi_{2,1}$, $\phi_{3,1}$,
  $\phi_{3,2}$ and $\theta_{2,1}$, $\theta_{3,1}$, $\theta_{3,2}$ compared
  with rate achieved using zero forcing precoding for
  $9\times3$ $(\mathrm{R}_{9\times3,\mathrm{ZF}})$, $6\times3$
  $(\mathrm{R}_{6\times3,\mathrm{ZF}})$ and $3\times3$ $(\mathrm{
    R}_{3\times3,\mathrm{ZF}})$ systems, where $\mathrm{C}_{9\times3}$, $\mathrm{
    C}_{6\times3}$ and $\mathrm{C}_{3\times3}$ are the respective system
  capacities at $10$ dB and $\mathrm{R}_{3\times3,\mathrm{si}}$ is obtained by using  only the best eigen mode in a $3\times3$ system.}
\end{figure}
For systems with $2$ antenna UEs, from Fig.~\ref{Rate_fig_mul}, we see
that the achievable rate is almost the downlink capacity (i.e. when
perfect CSI is available) if about $10$ bits are used for quantization of
the parameters $\phi$, $\theta$ for the $2\times 2$ post-coder
matrix. More significantly, when compared to the ZF
precoding based approach, we find that the significant gains in rate
can be observed when the post-coder is used with $4$ or more bits of
quantization. In the $2\times 2$ case, since the optimization to
perform transmit domain ZF precoding does not yield a feasible solution, we
selectively invert only the best channel as a comparison with the
feedback based approach. We find that even with just $3$ bits of
quantization, the performance using feedback far exceeds that achievable using purely
transmitter compensation, thereby justifying the use of post-coder
feedback.

For systems with $3$ antennas at the UE, the performance trends are
similar to those of systems with $2$ antennas, except that it requires a
larger number of bits to achieve rates close to the upperbound, as seen in
Fig.~\ref{lower_bound_mul_ant_3ue}. From %Fig.~\ref{fig:lloyd:upper}, one can
Observe that the effect of quantization becomes small after $24$ bits in Fig.~\ref{fig:lloyd:upper}. This manifests as enhanced achievable rates in Fig.~\ref{lower_bound_mul_ant_3ue}. %It must be
%mentioned that the high feedback requirement in terms of bits can be
%amortized by using a low rate %adaptive approach, such as the one
%mentioned in~\cite{roh2007efficient}, to minimize feedback
%requirements without affecting performance.
For the $3\times 3$ case,
we also find that selectively inverting to use only the best channel yields
a much poorer performance.
%this is marked as $R_{3\times3,si}$ in Fig.~\ref{lower_bound_mul_ant_3ue}.
%\sibi{remove(ignoring worst channel might improve the rate,
%which comes at additional cost of computing),}
Clearly, post-coder feedback is always better in these systems.
% \begin{figure}[!h]
% \centering
% \includegraphics[width=0.8\textwidth]{./figures/qe_err_3ant}
% \caption{\label{qe_err_3ant}$\mathbb{E}[\bQ]$ vs number of bits used for quantization for system with 3 receive antennas.}
% \end{figure}

\begin{figure}[!h]
\centering
\includegraphics[width=\textwidth]{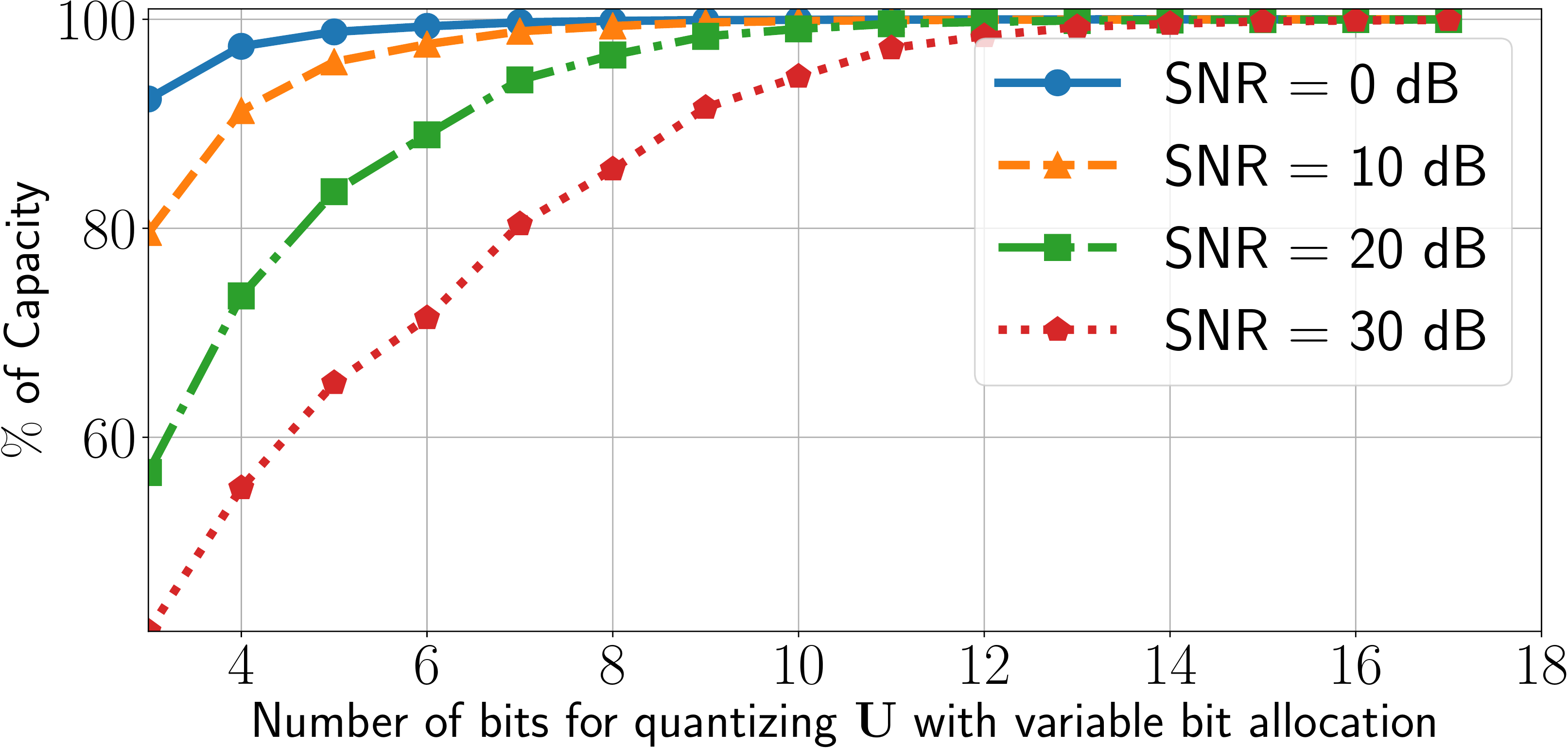}
\caption{\label{Rate_ratio_fig} Percentage of capacity achieved for a $4\times 2$ MIMO system at different SNRs.}
\end{figure}

\begin{comment}
Fig.~\ref{massiveMIMO} shows the
rate for a massive MIMO system with $1000$ transmit antennas and $2$
antennas at UE. There is a significant increase in
the achievable rate when compared to the case with a smaller
number of BS antennas, since the singular values of the channel are
significantly larger.
% TODO: In Fig. 3, the second label should be "Rate using equal bit
% allocation for $\bU$ parameters". Third label should be "Rate using variable bit
% allocation for $\bU$ parameters"

\begin{figure}[!h]
\centering
\includegraphics[width=0.8\textwidth]{./figures/asymptotic_bound}
\caption{\label{massiveMIMO} Lower bound on achievable rate for a massive MIMO system with $1000$ transmit antennas and $2$ receive antennas for equal bit allocation for both $\theta$ and $\phi_3$.}
\end{figure}

%Finally,
\end{comment}
The variation of the achievable rate with SNR for a $4\times 2$ MIMO
system is shown in Fig.~\ref{Rate_ratio_fig}, where the ratio of the
achievable rate using  $\hat\bU$ against the downlink
capacity is shown as a percentage. Observe that, as the SNR increases,
achieving a rate close to capacity requires a larger number of
bits. This is mainly because efficient allocation of resources at high
SNR requires a more accurate $\hat\bU$ at the receiver.

\subsection{Effective utilization of bits alloted}
As discussed in the previous section, when the quantization bit budget
is low, allotting bits to all the singular vectors of the channel may
not be effective for systems having more than $3$ antennas at UE,
since inaccurate quantization of singular vectors would reduce the
rate significantly.  An alternate approach is to allocate all power to
a single singular vector and quantize only that, since the benefit
from more accurate quantization could outweigh the benefit from using
all channels in this situation. Therefore, we also study the effect of
quantizing just one singular vector, rather than the full
matrix. Notice that our quantization scheme using Givens rotation
cannot directly yield individual quantizers for each of the singular
vectors. We, therefore, parameterize only the best singular vector
with all available bits to study the performance, as described in
Section~\ref{sec:downlinkrates}.

\begin{figure}[!h]
\centering
\includegraphics[width=0.8\textwidth]{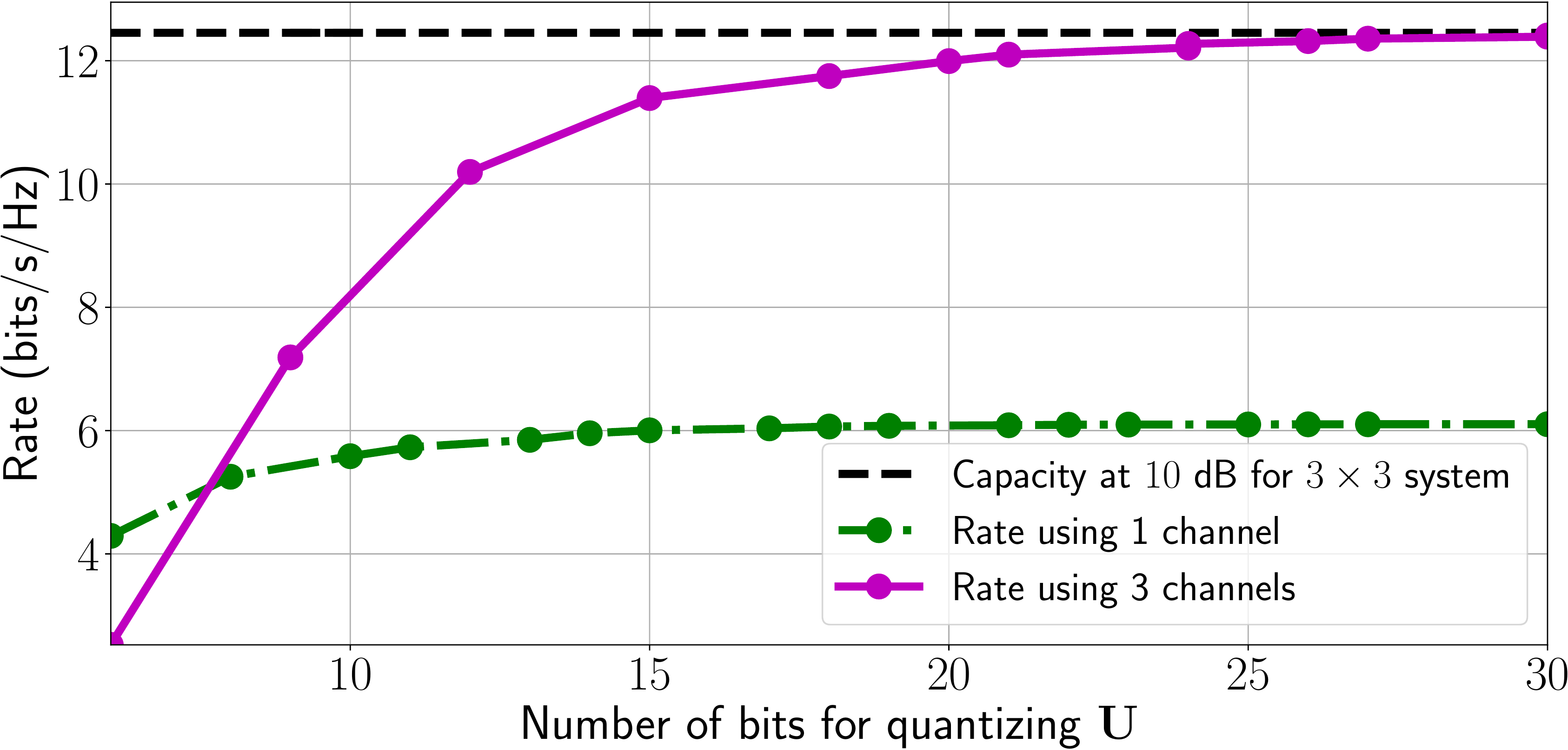}
\caption{\label{channel_fig} Rate achieved vs channels used, for $3\times3$ system.}
\end{figure}

In Fig.~\ref{channel_fig} we see that, when using fewer than $8$ bits
for quantizing the post-coder $\bU$, the rate achieved using a single
channel is higher than the rate achieved using three channels. This is
because, for one channel, the number of parameters required to be fed
back are just $4$ i.e, $\phi_{2,2}$, $\phi_{2,3}$, $\theta_{1,1}$ and
$\theta_{1,2}$, whereas 6 parameters are needed for the full matrix,
viz. $\phi_{2,2}$, $\phi_{2,3}$, $\phi_{3,3}$, $\theta_{1,1}$,
$\theta_{1,2}$ and $\theta_{2,1}$. Therefore, when the number bits is
small, quantizing only the best singular vector's parameters results
in more useful CSI than when quantizing all channel parameters
simultaneously with the same bit budget. This benefit diminishes as
the number of bits used for quantization increases, and the rate
achieved using one channel saturates, while using three channels
yields better performance.

% \sibi{ The below looks superfluous, can we remove?
% For systems with less than $3$ antennas at
% the UE, we have observed that there is no performance gain when a
% smaller number of bits is available for quantization of $\bU$. This
% can be attributed to the fact that the number of parameters required
% to represent the complete CSI matrix ($\bU$) is the same as the number
% of parameters required to represent the CSI matrix with only one
% singular value for a system with $2$ antennas at the UE (i.e., the first
% column of $\bU$).}

% \begin{figure}[!h]
% \centering
% \includegraphics[width=0.8\textwidth]{./figures/3channel_vs_rate1000_equal_allocation}
% \caption{\label{channel_fig1000} Rate vs channels for $1000\times3$ system with equal bit allocation for all parameters.}
% \end{figure}

\subsection{Bit error rate}
\begin{figure}[!h]
\centering
\includegraphics[width=1\textwidth]{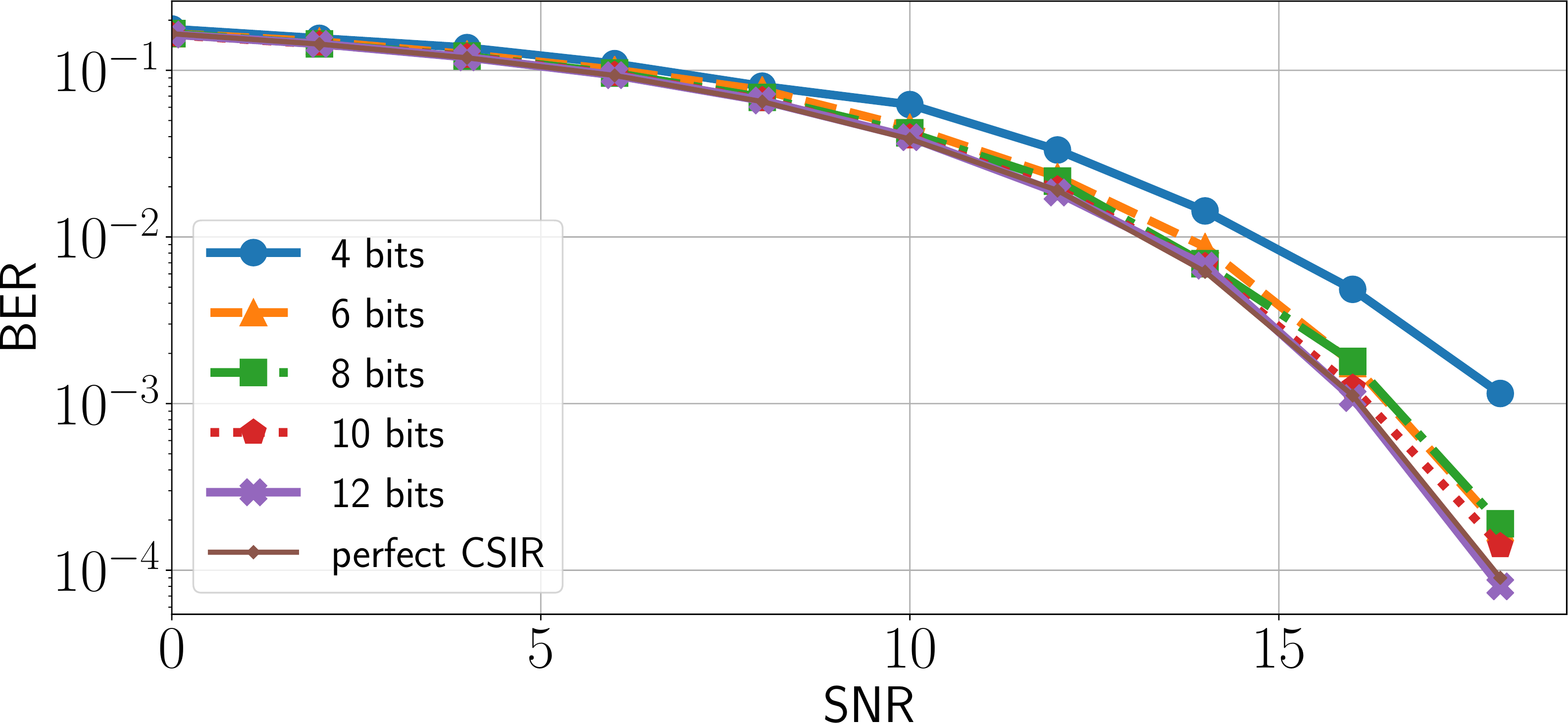}
\caption{\label{ber_fig} BER of QPSK with variable bit allocation for $\theta$ and $\phi$, of
  a $4\times2$ system at $10$dB.}
\end{figure}

In addition to high achievable rates, %effective parallelization of the
%channel parameters would also ensure that the
we now show that our  communication schemes maintain a low BER
%BER is kept low and does
on each parallel eigen mode of the MIMO channel. Thus the adverse effects due to the inaccurate post-coder are kept low.
%Let us consider
%the BER of the best channel %(eigenmode) for illustration.
In Fig.~\ref{ber_fig}, we observe that the BER obtained
when transmitting a QPSK signal over the spatial channel that
corresponds to the larger singular value of a $4\times 2$ MIMO system
with post-coder feedback is sensitive to the number of bits used to
quantize $\bU$. We see that the use of fewer than $10$ bits results in poor performance, since the $\hat{\bU}$ at the receiver is not sufficiently
accurate to diagonalize the channel. However, with $12$ bits,
the performance is close to that obtained using the perfect post-coder.

Although we considered the case where the post-coder is fully estimated
only at the BS and fed back to the UE, one could also estimate the
post-coders via downlink pilots at the UE. However, in this case,
refining the estimate at the UE may require more resources.
%In such a
%situation, low rate feedback to the UE to adapt the precoder using
%very few bits, similar to the approach discussed
%in~\cite{roh2007efficient} can refine the estimate and effectively
%track the channel variations.
%
We further remark that an
additional reduction in the
quantization bit requirement is feasible by adaptive quantization
approaches, particularly for slowly varying channels, as described in~\cite{roh2007efficient}.
For example, using just $1$ bit per parameter with an adaptive tracking mechanism may
further reduce the bit budget. The convergence of  $\hat{\bU}$  to $\bU$ is still governed by the analysis presented here.

\section{Conclusion}
\label{sec:conclusion}
We have considered downlink transmission in MIMO-TDD systems where UEs
possess multiple antennas, where post-coder information is used to
enhance performance. Achieving the capacity and enabling channel
parallelization requires that the receiver know the right singular
vectors of the MIMO channel matrix. We take the approach of using the
Givens rotations and Householder transformations to parameterize the
unitary post-coders, which permits us to represent unitary matrices in
terms of independent scalar parameters that represent rotation
angles. When the UE has 2 antennas, we show that about $8$ bits
suffice to quantize the post-coder accurately. This requirement can
be reduced to less than 2 bits per training instant when adaptive
tracking of the parameters is employed. When the UE has 3 antennas 26
bits are sufficient to accurately represent the post-coder, with
adaptive refinement requiring only 6 bits per instant. We further show
that, when fewer bits are available, quantizing the dominant singular
vector alone achieves higher rates than quantizing and feeding back
the complete post-coder. Our simulations reveal that the proposed
quantization approach yields performance close to that achieved using
the accurate post-coder, both in terms of being close to the link
capacity as well as achieving low BER. Future work would focus on
extending these techniques to UEs to multi-user scenarios where
post-coder feedback can be used for eliminating interference and
aligning transmissions for various users.

\renewcommand{\bibfont}{\footnotesize}
\bibliographystyle{IEEEtran} \bibliography{IEEEfull,references}

\begin{appendices}

\section{Proof of Theorem~\ref{thm:general}
  %matrix $\bQ$
  }\label{upper_experr}

%Using Equation~\eqref{eq:pc:mat:rep}, the post-coder matrix $\bU$ can be written as
% \begin{equation}
%   \bU=\left[\prod_{k=1}^{r}\bD_{k}(\phi_{k, 1}, \ldots, \phi_{k, k})\prod_{l=1}^{r-k}\bG_{r-l, r-l+1}(\theta_{k, l})\right]\bI.
% \end{equation}
Denoting $\bD_{k}(\phi_{k, 1}, \ldots, \phi_{k, k})$ in \eqref{eq:pc:mat:rep:2} as $\bD_k$, and $\bG_{r-l,
  r-l+1}(\theta_{k, l})$ as  $\bG_{r-l}^{k,l}$, we have
 \begin{equation*}
   \bU=\left[\prod_{k=1}^{r-1}\bD_{k}\prod_{l=1}^{r-k}\bG_{r-l}^{k,l}\right]\bI.
 \end{equation*}
 Similarly, the quantized post-coder matrix $\hat\bU$ is given as
 \begin{equation*}
   \hat\bU=\left[\prod_{k=1}^{r-1}\hat\bD_{k}\prod_{l=1}^{r-k}\hat\bG_{r-l}^{k,l}\right]\bI,
 \end{equation*}
where $\bD_{k}(\hat\phi_{k, 1}, \cdots, \hat\phi_{k, k}) = \hat\bD_k$ and $\bG_{r-l, r-l+1}(\hat\theta_{r, l})=\hat\bG_{r-l}^{k,l}$. Letting
$\bQ=(\hat{\bU}^{\dagger}\bU - \bI_r)^{\dagger}(\hat{\bU}^{\dagger}\bU
- \bI_r)$,
\begin{equation*}
  \Tr(\bQ) = \norm{\bU-\hat\bU}^2_F
      = \norm{\left[\prod_{k=1}^{r-1}\bD_{k}\prod_{l=1}^{r-k}\bG_{r-l}^{k,l}\right]-\left[\prod_{k=1}^{r-1}\hat\bD_{k}\prod_{l=1}^{r-k}\hat\bG_{r-l}^{k,l}\right]}^2_F
\end{equation*}
Taking $\mathbf{E}_1=\bD_1-\hat\bD_1$, and
applying triangle inequality on the above expression
\begin{equation*}
  \begin{split}
    \Tr(\bQ) & \leq \norm{\left[\prod_{l=1}^{r-1}\bG_{r-l}^{1,l}\prod_{k=2}^{r-1}\bD_{k}\prod_{l=1}^{r-k}\bG_{r-l}^{k,l}\right]-\left[\prod_{l=1}^{r-1}\hat\bG_{r-l}^{1,l}\prod_{k=2}^{r-1}\hat\bD_{k}\prod_{l=1}^{r-k}\hat\bG_{r-l}^{k,l}\right]}^2_F+\norm{\mathbf{E}_1\bD_1\hat{\bD}_1}^2_F\\
    & = \norm{\left[\prod_{l=1}^{r-1}\bG_{r-l}^{1,l}\prod_{k=2}^{r-1}\bD_{k}\prod_{l=1}^{r-k}\bG_{r-l}^{k,l}\right]-\left[\prod_{l=1}^{r-1}\hat\bG_{r-l}^{1,l}\prod_{k=2}^{r-1}\hat\bD_{k}\prod_{l=1}^{r-k}\hat\bG_{r-l}^{k,l}\right]}^2_F+\norm{\mathbf{E}_1}^2_F.
  \end{split}
\end{equation*}
Now, taking $\mathbf{E}^{'}_{1,r-1}=\bG_{1}^{1,r-1}-\hat\bG_{1}^{1,r-1}$,
\begin{equation*}
  \begin{split}
    \Tr(\bQ) & \leq \norm{\left[\prod_{l=1}^{r-2}\bG_{r-l}^{1,l}\prod_{k=2}^{r-1}\bD_{k}\prod_{l=1}^{r-k}\bG_{r-l}^{k,l}\right]-\left[\prod_{l=1}^{r-2}\hat\bG_{r-l}^{1,l}\prod_{k=2}^{r-1}\hat\bD_{k}\prod_{l=1}^{r-k}\hat\bG_{r-l}^{k,l}\right]}^2_F+\norm{\mathbf{E}_1}^2_F+\norm{\mathbf{E}^{'}_{1,r-1}\bG_{1}^{1,r-1}\hat{\bG}_{1}^{1,r-1}}^2_F\\
    & = \norm{\left[\prod_{l=1}^{r-2}\bG_{r-l}^{1,l}\prod_{k=2}^{r-1}\bD_{k}\prod_{l=1}^{r-k}\bG_{r-l}^{k,l}\right]-\left[\prod_{l=1}^{r-2}\hat\bG_{r-l}^{1,l}\prod_{k=2}^{r-1}\hat\bD_{k}\prod_{l=1}^{r-k}\hat\bG_{r-l}^{k,l}\right]}^2_F+\norm{\mathbf{E}_1}^2_F+\norm{\mathbf{E}^{'}_{1,r-1}}^2_F
  \end{split}
\end{equation*}
%where
%using the same approach we can reduce the above equation to
We now observe that successive application of the triangle inequality
yields norms of matrices of the form $\bE_i$ and $\bE_{k,l}'$. These
are easy to characterize. $\bE_i$ is a diagonal matrix
whose diagonal entries consist of $r-k$ $1$s followed by terms
$e^{\phi_{k,k}} - e^{\hat{\phi}_{k,k}}, e^{\phi_{k,+1}} -
e^{\hat{\phi}_{k,k+1}} \ldots e^{\phi_{k,r}} -
e^{\hat{\phi}_{k,r}}$. The $\bE_{k,l}'$ matrix has all zeros except
for four entries, where two of them are $\cos\theta_{k,l} -
\cos\hat{\theta}_{k,l}$, and the other two $\sin\theta_{k,l} - \sin\hat{\theta}_{k,l}$.
Using this, on further simplification, we get
\begin{equation}\label{eq:trace:bnd}
    \Tr(\bQ)  \leq \sum_{i=1}^{r-1}\norm{\mathbf{E}_i}^2_F+\sum_{k=1}^{r-1}\sum_{l=1}^{r-k}\norm{\mathbf{E}^{'}_{k,l}}^2_F
\end{equation}
where $\mathbf{E}_i = \bD_i-\hat\bD_i$ and $\mathbf{E}^{'}_{k,l} =
\bG_{r-l}^{k,l}-\hat\bG_{r-l}^{k,l}$. We now note that each of
%\begin{equation}\label{qbound}
%    \mathbb{E}[\Tr(\bQ)]  \leq %\sum_{i=1}^{r}\mathbb{E}\left[\norm{\mathbf{E}_i}^2_F\right]+\sum_{k=1}^{r-1}\sum_{l=1}^{r-k}\mathbb{E}\left[\norm{\mathbf{E}^{'}_{k,l}}^2_F\right]
%\end{equation}
\begin{equation*}
  \begin{split}
    \mathbb{E}\left[\norm{\mathbf{E}_i}^2_F\right]& = \mathbb{E}\left[\Tr(\bD_i-\hat\bD_i)(\bD_i-\hat\bD_i)^{\dagger}\right]\\
    & = \mathbb{E}\left[\sum_{l=i}^{r}\norm{e^{j\phi_{i,l}}-e^{j\hat\phi_{i,l}}}^2\right]\\
    & = \sum_{l=i}^{r}\mathbb{E}\left[\norm{e^{j\phi_{i,l}}-e^{j\hat\phi_{i,l}}}^2\right].
  \end{split}
\end{equation*}
Under uniform quantization of $\phi_{i,j}$
\begin{equation*}
  \mathbb{E}\left[\norm{e^{j\phi_{i,l}}-e^{j\hat\phi_{i,l}}}^2\right] = 2-2\mbox{sinc}(2^{-b_1}), \: \forall \: l = 1 \ldots r, \: i={l\ldots r-1}.
\end{equation*}
Thus, we get
\begin{equation}\label{expec}
    \mathbb{E}\left[\norm{\mathbf{E}_i}^2_F\right] = 2r(1-\mbox{sinc}(2^{-b_1})).
\end{equation}
Now $
    \mathbb{E}\left[\norm{\mathbf{E}^{'}_{k,l}}^2_F\right]  = \mathbb{E}\left[\norm{\bG_{i}^{k,l}-\hat\bG_{i}^{k,l}}^2_F\right]$ can be evaluated as
\begin{equation}\label{expectheta}
  \begin{split}
    \mathbb{E}\left[\norm{\mathbf{E}^{'}_{k,l}}^2_F\right] &= 2\mathbb{E}\left[\norm{\cos\theta_{k,l}-\cos\hat\theta_{k,l}}^2+\norm{\sin\theta_{k,l}-\sin\hat\theta_{k,l}}^2\right]\\
    & = 4\mathbb{E}[1-\cos(\theta_{k,l}-\hat\theta_{k,l})]\\
    & = 8 \mathbb{E}\left[\sin\left(\frac{\theta_{k,l}-\hat\theta_{k,l}}{2}\right)^2\right].
    \end{split}
\end{equation}
Then, from~\eqref{eq:trace:bnd}, ~\eqref{expec} and ~\eqref{expectheta}, we get
\begin{equation}\label{eq:trace:expect}
  \mathbb{E}[\Tr(\bQ)]  = \epsilon_{\mathrm{Lloyd}}^2 \leq r(r-1)(1-\mbox{sinc}(2^{-b_1}))+\sum_{k=1}^{r-1}\sum_{l=1}^{r-k}8\left(\mathbb{E}\left[\sin\left(\frac{\theta_{k,l}-\hat\theta_{k,l}}{2}\right)^2\right]\right).
\end{equation}
%In the above equation the first term contains \mbox{sinc}($2^{-b_1}$)$\to 0 $ and the second term also approaches $0$ as the number of bits ($b_2(l)$) for quantization increases, there by reducing the quantization error to 0.
%Further, if $\theta_{k,l} \approx \hat\theta_{k,l}$ then $\mathbb{E}\left[\sin\left(\frac{\theta_{k,l}-\hat\theta_{k,l}}{2}\right)^2\right]$ can be reduced as follows,
Since
\begin{equation*}
  \frac{\theta_{k,l}-\hat\theta_{k,l}}{2} \geq \sin\left(\frac{\theta_{k,l}-\hat\theta_{k,l}}{2}\right),
\end{equation*}
the MSE $\stl$ for $\theta_{k,l}$ can be bounded as
%be represented as $\tilde{\theta}^2_{k,l}$. Then
\begin{equation}\label{eq:bound:stl}
\stl= \mathbb{E}\left[(\theta_{k,l} - \hat{\theta}_{k,l})^2\right] \geq 4\mathbb{E}\left[\sin\left(\frac{\theta_{k,l}-\hat\theta_{k,l}}{2}\right)\right]^2.
\end{equation}
Recall that the density function
of $\theta_{k,l}$ is independent of $k$, from Equation~\eqref{eq:theta:pdf}.
%Thus, $\tilde{\theta}^2_{k,l}$ depends only on $l$.
Combining \eqref{eq:trace:expect} and \eqref{eq:bound:stl} completes the proof of the theorem.

\section{Expected Post-coder Quantization Error for $r=2$}\label{expecq}

In the following proof, we consider the decomposition for the precoder to be similar to the one in~\cite{roh2007efficient}, which involves the form that we have used in Section~\ref{code_book}.

Denoting the post-coder quantization error $\bQ=({\bU}^{\dagger}\hat{\bU} - \bI_r)^{\dagger}({\bU}^{\dagger}\hat{\bU}
- \bI_r)$  by the matrix $\bQ$,
%\begin{equation*}
 % \bQ = (\hat{\bU}^\dagger\bU-\bI_r)^\dagger
%(\hat{\bU}^\dagger\bU-\bI_r)
%\end{equation*}
\begin{equation}\label{eq:12}
  \mathbb{E}[\bQ] = 2\bI_r - \mathbb{E}[{\bU}^{\dagger}\hat{\bU}] - \mathbb{E}[\hat{\bU}^{\dagger} \bU]
\end{equation}
To evaluate the above, we use the fact that all $2\times 2$ unitary
matrices can be represented using Givens rotations and Householder
transformations as follows (with the final diagonal unitary matrix not
represented due to redundancy, as discussed shown in Equation~\eqref{eq:pc:mat:rep:2}):
\begin{equation}\label{eq:10}
  {\bU} =
  \underbrace{\begin{bmatrix}
      1& 0 &\\
      0&  e^{j \phi_1}&
  \end{bmatrix}}_{{\bU}_1}
  \underbrace{\begin{bmatrix}
      \cos  \theta & \sin  \theta &\\
      -\sin  \theta & \cos  \theta &
  \end{bmatrix}}_{{\bU}_2}
  % \underbrace{\begin{bmatrix}
  %     e^{j \phi_2}& 0 &\\
  %     0&  e^{j \phi_3}&
  % \end{bmatrix}}_{{\bU}_3}
.
\end{equation}
Note that we refer to the component matrices as $\bU_1$ and $\bU_2$ as marked
in the equation~\eqref{eq:10} for convenience. Therefore, $\phi_1$, and
$\theta$ uniquely parameterize all $2\times 2$ unitary
post-coders. Since we use scalar quantization to quantize these
parameters, the matrix $\hat{\bU}$ obtained by reconstruction using
the corresponding quantized $\hat{\phi}_1$ and $\hat\theta$ is
\begin{equation}\label{eq:11}
  \hat{\bU} =
  \underbrace{\begin{bmatrix}
	    1& 0 &\\
				0&  e^{j\hat \phi_1}&
	\end{bmatrix}}_{\hat{\bU}_1}
	\underbrace{\begin{bmatrix}
	    \cos \hat \theta & \sin \hat \theta &\\
	    -\sin \hat \theta & \cos \hat \theta &
	\end{bmatrix}}_{\hat{\bU}_2}
  %   \underbrace{\begin{bmatrix}
  %     e^{j\hat \phi_1}& 0 &\\
  %     0&  e^{j\hat \phi_2}&
  % \end{bmatrix}}_{\hat{\bU}_3}
.
\end{equation}
We can use the component matrices $\bU_1$ and $\bU_2$ in
Equation~\eqref{eq:10} and $\hat{\bU}_1$ and $\hat{\bU}_2$ in
Equation~\eqref{eq:11} to evaluate the expectation in
Equation~\eqref{eq:12}. This allows us to exploit the independence of
the parameters in computing the expectation. We first consider the
matrix
${\bU}^\dagger\hat{\bU}
={\bU}^\dagger_2{\bU}^\dagger_1\hat{\bU}_1\hat{\bU}_2$. Since
$\hat\bU_1^\dagger\bU_1$ depends only on $\phi_1$, this
expectation can be evaluated separately as
\begin{equation}\label{u1hatu1}
  \mathbb{E}\left[{\bU}_1^\dagger\hat{\bU}_1\right] = \begin{bmatrix}
				1 & 0 \\
				0 & \mathbb{E}[e^{j(\hat{\phi}_1 - \phi_1)}].
  \end{bmatrix}
\end{equation}
If we use a uniform $b$-bit quantizer for each $\phi_1$ that has
$2^b$ equi-spaced levels in $(-\pi, \pi]$, then we obtain $\mathbb{E}\left[{\bU}_1^\dagger\hat{\bU}_1\right]
= \mbox{diag}\{1, \mbox{sinc}(2^{-b})\}$. This
simplifies the evaluation of the expectation of $\bQ$, since, using
the rule of iterated expectations and that $\mathbb{E}[\bQ] =   \mathbb{E}\left[{\bU}^\dagger_2\mathbb{E}[{\bU}^\dagger_1\hat{\bU}_1\vert\bU_2
      ]\hat{\bU}_2\right] $, we get
\begin{equation*}
\mathbb{E}[\bQ]= \begin{bmatrix}
	  \mathbb{E}[\cos(\theta)\cos(\hat{\theta}) + \sin(\theta)\sin(\hat{\theta})\mbox{sinc}(2^{-b})] & \mathbb{E}[\cos(\hat{\theta})\sin(\theta) - \cos(\theta)\sin(\hat{\theta})\mbox{sinc}(2^{-b})]\\
      \mathbb{E}[\cos(\theta)\sin(\hat{\theta}) - \cos(\hat{\theta})\sin(\theta)
      \mbox{sinc}(2^{-b})] & \mathbb{E}[\sin(\theta)\sin(\hat{\theta}) + \cos(\theta)\cos(\hat{\theta})\mbox{sinc}(2^{-b})]
\end{bmatrix}
\end{equation*}
Therefore, substituting this into Equation~\eqref{eq:12}, and taking
trace, we get
  \begin{equation*}\label{eq:24}
\epsilon_{\mathrm{Lloyd}}^2=    \mathbb{E}[\mbox{Tr}(\bQ)] = 4 - 2\mathbb{E}\left[\cos(\theta - \hat{\theta})\right](1 + \mbox{sinc}(2^{-b}))
  \end{equation*}
which is the desired result.
%   In the above, $C$ can be evaluated by integration in
%   Equation~\eqref{eq:theta:pdf}, by splitting the integral over intervals
%   that correspond to the quantized value of $\theta$. While a closed
%   form does not exist, this is simple to approximate using trapezoidal
%   numerical integration. The entries of the above matrix are real. So,
%   since
%   $\mathbb{E}[\hat\bU^\dagger \bU] = \mathbb{E}[\bU^\dagger\hat\bU]$
%   in Equation~\eqref{eq:12}, we get
% \begin{equation*}
%   \mathbb{E}[\bQ]= 2\mbox{sinc}(2^{-b})\begin{bmatrix}
%     1-C & 0 \\
% 	 0 & 1-\mbox{sinc}(2^{-b}) C
% \end{bmatrix}.
% \end{equation*}
% Taking trace of the above yields the desired result, that is:
% \begin{equation*}
% \epsilon_{\mathrm{Lloyd}}^2 = 2\mathrm{sinc}(2^{-b})\left(2 - (1 + \mathrm{sinc}(2^{-b}))
%   \mathbb{E}\left[\cos(\theta - \hat{\theta})
% \right]
% \right).
% \end{equation*}
\section{Lower bound on Achievable rates}\label{lbproof}
Expressing ${\bf H} = \bU \Sigma \bV^{\dagger}$ by SVD, and taking  $\tilde{\bx} = \bV \bx$, we get
\begin{equation*}
  \tilde{\by} = \bU \Sigma \bV^{\dagger}\tilde{\bx} + \tilde{\eta} = \bU\Sigma \bx + \tet.
\end{equation*}
Let $y = \hat{\bU}^{\dagger} \tilde{\by}$, where $\hat{\bU}$ is the quantized version of $\bU$ available at the UE. We have
\begin{equation}\label{neqno:62}
  \by = \hat{\bU}^{\dagger}\bU\Sigma \bx + \hat{\bU}^{\dagger}\tilde{\eta} = \hat{\bU}^{\dagger}\bU\Sigma \bx + \bw,
\end{equation}
where $\bw =\hat{\bU}^{\dagger}\eta$. Notice that $\bw$ and $\tet$ are identically distributed, and independent of $(\bx,\bH)$.
We have assumed $\eta\sim \mathcal{NC}(0,\bI)$.
Now, instead of the optimization in  \eqref{eq:mut:inf}, one can equivalently
consider maximizing the mutual information $I(\bx;\by|\Sigma, \hat \bU)$,
under the constraint $\eE[\bx^{\dagger}\bx] \leq P_T$. As mentioned earlier,
the exact nature of the optimal distribution is unclear in the absence of full CSIR, and appears difficult to characterize.
Nevertheless, notice that $\Cfcsi$ is an upperbound to the achievable rate here, thus, we can focus on a lowerbound which is close enough to
$\Cfcsi$ itself.
In particular, the choice $\bx \sim \mathcal{CN}(0,\bK_{\Sigma})$, for an appropriate covariance matrix $\bK_{\Sigma}$, will be shown to achieve rates close to the capacity as the quantization gets finer. The achievable rate $R$ then is given by
\begin{align}
  \label{eq:1}
  R = I(\bx;\by|\Sigma, \hat \bU)
  &= h(\by|\Sigma, \hat \bU) - h(\by|\Sigma, \hat \bU, \bx) \\
  &\geq h(\by|\Sigma, \hat \bU, \bU) - h(\by|\Sigma, \hat \bU, \bx) \\
  &= h(\bU\Sigma \bx + \tet|\Sigma,\bU, \hat \bU) - h(\by|\Sigma, \hat \bU, \bx).
\end{align}
The inequality follows from the fact that conditioning reduces differential entropy. Notice that $(\Sigma, \bU, \hat \bU) \rightarrow \bU\Sigma \rightarrow \bU \Sigma x+\tet$ forms a Markov chain in this order. Thus,
\begin{align}
  R &\geq  h(\bU\Sigma \bx + \tet|\bU\Sigma) - h(\by|\Sigma, \hat \bU, \bx) \notag \\
  &= h(\bU\Sigma \bx + \tet|\bU\Sigma) - r \log (\pi e)  - \left[ h(\by|\Sigma, \hat \bU, \bx) - r \log (\pi e)\right]\notag \\
  &= \Rfcsi - h(\by|\Sigma, \hat \bU, \bx) + r \log (\pi e), \label{eq:rate:up:2}
\end{align}
where $\Rfcsi$ is the achievable rate when ${\bf H}$ is known fully at the transmitter and receiver, while taking $\bx \sim \mathcal{CN}(0,\bK_{\Sigma})$.
Let us now find a suitable upperbound to the term $h(\by|\Sigma, \hat \bU, \bx)$. Rewriting \eqref{neqno:62} we get,
\begin{align}
  h(\by|\Sigma, \hat \bU, \bx) &=
  h(\Sigma \bx + (\hat{\bU}^{\dagger}\bU - \bI_r)\Sigma \bx + \bw|\Sigma, \hat \bU, \bx) \notag \\
  &= h\left( (\hat{\bU}^{\dagger}\bU - \bI_r) \Sigma \bx + \bw|\Sigma, \hat \bU, \bx\right)\notag \\
  &\leq h(\hat{\bU}^{\dagger}\bU - \bI_r) \Sigma \bx + \bw).
\end{align}
%Defining $\bQ=(\hat{\bU}^{\dagger}\bU - \bI_r)(\hat{\bU}^{\dagger}\bU
%- \bI_r)^\dagger$,
Applying the entropy maximizing property of Gaussian distribution under a covariance constraint
\begin{align*}
  h(\by|\Sigma, \hat \bU, \bx) &\leq \eE\left[ \log |\pi e (\eQ\Sigma K_{\Sigma} \Sigma^{\dagger} {\eQ}^{\dagger} + N_o\bI_r)|\right].
\end{align*}
%Observe that the expectation above is over the fading distribution across blocks.
%In each block the proposed feedback scheme is used, and it results in a certain achievable rate. The quantity of interest is the average of these
%achievable rates $\aRate$, or the ergodic rate, which is indicative of the true throughput achieved by the system.
Let us denote $B = \hat\bU^{\dagger}\bU-\bI_r$. Since
$\det(\bI+\mathbf{A}\mathbf{D}) = \det(\bI+\mathbf{D}\mathbf{A})$ when  both $\mathbf{AD}$ and $\mathbf{DA}$ exist,
%%
%
%\begin{equation}
%  \label{eqn:lbound1}
%  h(\by|\Sigma, \hat \bU, \bx) \leq \mathbb{E}\left[\log_{2} \left\{\det
%    \left(\pi e\left( \bI_r+\bB\Sigma \bK_\Sigma
%      \Sigma^{\dagger}\bB^{\dagger}\right)\right)\right\}\right]
%\end{equation}
%For any appropriately sized matrices $\mathbf{A}$ and $\mathbf{D}$, we know that
%\begin{equation*}
%  \det(\bI+\mathbf{A}\mathbf{D}) = \det(\bI+\mathbf{D}\mathbf{A})
%\end{equation*}
%
%Using the above identity, we transform~\eqref{eqn:lbound1} to
\begin{equation*}
  h(\by|\Sigma, \hat \bU, \bx) \leq \mathbb{E}\left[\log_{2}
    \left\{\det \left( \pi e \left(\bI_r+\Sigma \bK_\Sigma
    \Sigma^{\dagger}\bB^{\dagger}\bB\right)\right)\right\}\right]
\end{equation*}
Notice that  $\bB^{\dagger}\bB = {\bf Q}$ is a matrix that captures
the error between $\bU$ and $\hat{\bU}$. Denoting $\bJ  = \Sigma \bK_{\Sigma} \Sigma^{\dagger}$,
%We refer to this as the
%post-coder error matrix. Thus,
%
\begin{align}
  h(\by|\Sigma, \hat \bU, \bx)
    &\leq \mathbb{E}\left[
    \log_{2} \left\{\det\left( \pi e \left(\bI_r+
    %\Sigma \bK_\Sigma \Sigma^{\dagger}
    \bJ \bQ\right)\right)
\right\}\right] \label{eq:qe1}  \\
    & \leq \mathbb{E}  \, \log_{2} \left\{ \prod_{i=1}^{r} \pi e \left(1+\bJ_{ii} \bQ_{ii} \right ) \right\} ,
%end{split}
%\end{equation}
\end{align}
by applying the Hadamard's inequality~\cite{garling2007inequalities}
%states that,
for positive semidefinite  matrices.  Now applying Jensen's inequality
\begin{align}\label{eq:8}
 h(\by|\Sigma, \hat \bU, \bx)
    & \leq  \sum_{i=1}^{r} \left[ \log_{2}\left(1+\mathbb{E}\left\{\bJ_{ii} \right\} \mathbb{E}\{\bQ_{ii}\} \right ) \right] + r \log (\pi e),
\end{align}
since the logarithm is a concave function.
%
%where the last inequality is due to Jensen's inequality, and the
%
Notice that the expectations over $\bJ_{ii}$ and $\bQ_{ii}$ can be separated since
they are independent ($\bJ$ depends only on $\Sigma$, while $\bQ$
depends only on $\bU$). From
\eqref{eq:rate:up:2} and \eqref{eq:8}
\begin{equation}
  \label{eq:tightlowerbound}
  \begin{split}
    I(\bx;\by |\Sigma) & \geq \Rfcsi - \sum_{ii=1}^{r} \left[ \log_{2}\left(1+\mathbb{E}\left\{\bJ_{ii} \right\} \mathbb{E}\{\bQ_{ii}\} \right ) \right]. \\
  \end{split}
\end{equation}
Let us take the covariance of $\bx$ as  $\bK_\Sigma = \mbox{diag}\left[P_1, \cdots, P_r\right]$, with
\begin{equation}
  P_n = \left(\frac{1}{\lambda}-\frac{1}{\sigma_n^2}\right)^+, \label{eq:cfcsi:pow}
\end{equation}
where $\lambda$ is determined by \eqref{expectationpower}, and
$(a)^+ \coloneqq \max\{0,a\}$. Using \eqref{eq:pow:rx}, we get
$\eE \bigl[ \bJ_{ii} \bigr]  = \Prx, 1 \leq i \leq r$.
%is the power allocated in $i^{th}$ channel, So
%\begin{equation*}
%  \bJ_{ii}\leq P_{\mathrm{exp}} \: \forall \: ii = 1\ldots r
%\end{equation*}
%
Thus
\begin{align}
    \sum_{i=1}^{r}  \log_{2} \left( 1+ \eE \left\{\bJ_{ii} \right\}
    \eE\{\bQ_{ii}\} \right )
    =  \sum_{i=1}^{r}  \log_{2}\left(1 +\Prx \eE \bQ_{ii}  \right).
 \end{align}
Now, Jensen's inequality for  logsum implies that
%the above inequality can be written as
\begin{align}
    \sum_{i=1}^{r}  \log_{2}\left(1+ \Prx \mathbb{E}\{\bQ_{ii}\} \right )  & \leq  r \log_{2}\left(\frac{1}{r}\sum_{i=1}^r (1+\Prx \mathbb{E}\{\bQ_{ii})\} \right ) \\
    &= r\log_2\left(1+\frac{\Prx}{r}\eE \Tr(\bQ)\right).
\end{align}
%
%where $P_{\mathrm{exp}}$ is computed using~\eqref{expectationpower}.
Since $\Tr(\bQ) = ||\bU - \hat\bU||_F^2$, and $\Rfcsi=\Cfcsi$ under  \eqref{eq:cfcsi:pow}, the rate expression  in \eqref{eq:tightlowerbound} yields
\begin{equation}
  \label{looselowerbound}
  \begin{split}
    I(\bx;\by |\Sigma) & \geq C_{\text{full CSI}} - r \log_{2}\left(1+\frac{\Prx}{r}\mathbb{E} \norm{\bU-\hat\bU}_{F}^2 \right). \\
  \end{split}
\end{equation}
This completes the proof of Theorem~\ref{thm:lowerbound}.

\end{appendices}
\end{document}